# Statistical Survey of Chemical and Geometric Patterns on Protein Surfaces as a Blueprint for Protein-mimicking Nanoparticles

John M. McBride[1,*], Aleksei Koshevarnikov[3], Marta Siek[1], Bartosz A. Grzybowski[1,2,3,*], Tsvi Tlusty[1,2,*]

[1]*Center for Algorithmic and Robotized Synthesis, Institute for Basic Science, Ulsan 44919, South Korea*
[2]*Departments of Physics and Chemistry, UNIST, Ulsan 44919, South Korea*
[3]*Institute of Organic Chemistry, Polish Academy of Sciences, Warsaw 01-224, Poland*

[*]email: jmmcbride@protonmail.com;   nanogrzybowski@gmail.com;   tsvitlusty@gmail.com

Despite recent breakthroughs in understanding how protein sequence relates to structure and function, considerably less attention has been paid to the general features of protein surfaces beyond those regions involved in binding and catalysis. This paper provides a systematic survey of the universe of protein surfaces and quantifies the sizes, shapes, and curvatures of the positively/negatively charged and hydrophobic/hydrophilic surface patches as well as correlations between such patches. It then compares these statistics with the metrics characterizing nanoparticles functionalized with ligands terminated with positively and negatively charged ligands. These particles are of particular interest because they are also surface-patchy and have been shown to exhibit both antibiotic and anticancer activities – via selective interactions against various cellular structures – prompting loose analogies to proteins. Our analyses support such analogies in several respects (e.g., patterns of charged protrusions and hydrophobic niches similar to those observed in proteins), although there are also significant differences. Looking forward, this work provides a blueprint for the rational design of synthetic nanoobjects with further enhanced mimicry of proteins' surface properties.

## Introduction

Sequence,[1–3] structure,[4,5] dynamics,[6,7] and function[8] of proteins have been studied for decades and in great detail, culminating in recent breakthroughs in structure prediction[9–12] and *de novo* protein design.[13] In parallel, numerous works have studied protein surfaces, primarily the specific surface regions related to interactions (protein-protein interfaces, binding sites, and active sites).[14–23] More recently, growing attention has been given to the often expansive surface regions involved in nonspecific interactions (hub proteins,[24–] [26] liquid-liquid phase separation,[27,28] subcellular location,[29,30] solubility/aggregation propensity).[31–33] These regions are crucial in shaping the protein '*quinary structure*', which determines the statistical interactions with the multitude of molecular counterparts in the crowded milieu of living cells.[34–37] These vast networks of interactions, specific and mostly nonspecific and transient, govern how proteins efficiently navigate the vast space of possible interaction partners as they search for the correct targets.[23,38–40] Thus, the quinary structure of a protein may determine whether it will be rejected by the immune system or may induce deleterious interactions by binding to the wrong targets.[41] Despite the breadth of past work, the majority of studies on the statistics of protein surfaces have either had (i) large samples but a narrow focus, often limited to predicting protein-protein interfaces and drug binding [16–20] or (ii) a more general view, but small sample sizes and a limited range of analyses.[42–50] Some basic features of protein surfaces are generally acknowledged: hydropathy of surfaces is tailored to match the intended environment;[29] protein-protein interfaces are enriched in hydrophobic and aromatic residues compared to the rest of the surface;[16] hydrophobic and electrostatic patches are limited in size to avoid aggregation.[33,51] Surprisingly, however, we lack a comprehensive overview of the statistics of protein surfaces. Here, we provide such a survey of the geometry and chemistry of protein surfaces. To this end, we present a library of statistics and cross-correlations of surface properties and their spatial distribution on protein surfaces.

One of our main objectives has been to use this statistical description as a blueprint for producing protein-like ligand shells around nanoparticles. We have long been interested in mixed charge nanoparticles (MCNPs) [52–57] decorated with both negatively and positively charged ligands. We used such (+/-) ligand shells to engineer interparticle potentials for pH-selective precipitation[53] as well as self-assembly of nanoparticle supra-crystals.[55] Moreover, we showed that depending on the ratios of (+) to (-) ligands, MCNPs can be made to interact selectively with the walls of either Gram-positive or Gram-negative bacteria[56] and that they can selectively kill cancerous mammalian cells of various types while not harming non-cancerous cells.[53,57] Because these particles had sizes (~3-12 nm diameter) similar to proteins and presented (+/-) charge mosaics on their surfaces, we have long speculated that, at least surface-wise, they can be construed as loosely analogous to proteins – though, we lacked any quantitative evidence for such a statement.

In this context, we now compare proteins with MCNP surfaces simulated using Molecular Dynamics (MD). These simulations indicate that despite differences in molecular topology, both these surfaces mainly consist of charged, convex protrusions and hydrophobic, concave cavities. We detail the various similarities and differences to proteins and discuss how relatively straightforward modifications to the ligand linker and head group could lead to NPs that are more similar to and compatible with proteins. Some of the practical questions we pose are: (i) How similar are the sizes of the charged and hydrophobic patches? (ii) Should the charged moieties on the MCNPs be based on single charged groups or perhaps be terminated in di- or even tri-carboxylic acids (−) or quaternary ammonium salts (+)? (iii) How does the length of the alkyl chains influence the proportions of hydrophobic regions exposed to the solvent?

In a broader context, we suggest that translating similar analyses to other classes of synthetic nanoparticles will open paths to designing protein-mimicking coatings with tailored and well-controlled interactions with the biological environment, a critical requirement for emerging nanomedicine applications.

## Methods

**Protein surface analysis.**

To get a general understanding of the statistics of protein surfaces, we study a representative, nonredundant set of 14,963 protein structures from the protein data bank (PDB).[4] To this end, we adapt and combine a set of established software tools and a pre-processing pipeline.[22] In this pipeline, we first add hydrogens to PDB structures using Reduce[58] and add charges using PDB2PQR.[59] We extract the solvent-accessible surface using MSMS,[60] using a probe radius of 1.5 Å, equivalent to the surface that a water molecule may feel (**Figure 1**). The resulting surface is hereafter considered the effective definition of a protein surface. Next, the surface triangulation is regularized using PyMesh[61] so that most edge lengths are between 1-3 Å. We evaluate the electric potential on the surface mesh using APBS.[62] An empirical hydrogen-bonding potential is used to assign hydrogen-bonding scores to vertices, based on the nearest atoms and their bond angles.[63] Each surface point is labeled according to its closest amino acid and is assigned a hydrophobicity of that amino acid (Kyte-Doolittle).[64] Note that this effective hydrophobicity score is correlational – in reality, hydrophobicity is dictated by the interplay of geometry and chemistry[65], which would require running MD for each protein. However, the effective measure suffices[66] for our purpose.

The original sample of proteins had 16,200 in total, which were selected from a nonredundant set of proteins available in the PDB, corresponding exactly to natural sequences without signal peptides.[67] However, only 14,963 made it through the pipeline due to the inability of PDB2PQR to handle structures with too many disordered residues and the lack of hydrophobicity information for non-canonical and modified amino acids. Functional annotations are obtained by matching Gene Ontology terms via UniProt, using the Python package *GOATOOLS*.[8,68,69] NCBI Taxa annotations are obtained from UniProt, using the Python package ETE 3.[70]

**Curvature**

We calculate Gaussian and mean curvature at each point using the Python *trimesh* module, using a sphere radius of 1 Å.[71] Mean curvature distinguishes concave (negative mean curvature), flat (approximately zero), and convex (positive) regions. Gaussian curvature distinguishes saddle points (negative Gaussian curvature) from mostly convex or concave regions (positive). We calculate geodesic distances on the surface graph using the Dijkstra algorithm,[72] where edge weights are the Euclidean distance between vertices.

**Patchiness**

To study surface patchiness, we divide the surface into clusters of connected regions with similar surface properties. The data for each surface property are divided into regions of high, low, and medium values.

The boundaries between the regions are positioned so that each of the three groups includes about a third of the surface (note that we cannot achieve perfect equipartition for surface properties with skewed distributions; see **SI Fig. 1**). The protein surface is then divided into clusters by labeling each face in the surface triangulation with its property group label (mean of its vertices). Two triangles are within the same cluster if they are adjacent and have the same group label.

To quantify the shape of surface clusters, we use a simple topological measure $D$ termed "discosity," where $D \rightarrow 0$ is for elongated clusters and $D \rightarrow 1$ for disc-like clusters. The cluster perimeter $P_0$, the number of edges that are not shared, is calculated as $P_0 = 3F - S$, where $S$ is the number of shared edges in the cluster, and $F$ is the number of faces. The discosity is then normalized to range from 0 to 1, $D = (P_{max} - P_0)/(P_{max} - P_{min})$, where $P_{max}$ $F + 2$ is the perimeter of an elongated, linear cluster, and $P_{min} = (6F)^{1/2}$ is the perimeter of a hexagonal cluster. This normalization procedure assumes equilateral triangles. However, the triangulations used in this study are only approximately equilateral, which leads to a rare number of values lying beyond the $0-1$ range. Also, due to the spherical geometry, the $P_{max}$ is only correct up to a surface fraction of about 0.5 since it is impossible to create a strictly linear chain of triangles on a sphere above this surface fraction. For cluster sizes covering an area fraction greater than about 0.3, the discosity parameter does not mean exactly disc-like, but rather that it is not elongated and thin.

**Protein Simulations**

We perform molecular dynamics (MD) simulations[73] using GROMACS (version 2022.4)[74] of three monomeric proteins in water: guanylate kinase (*mycobacterium tuberculosis*, PDB ID 1ZNW), nirD (*mycobacterium tuberculosis*, PDB ID 4AIV), and cutinase (*hypocrea jecorina*, PDB ID 4PSC). We use a NaCl concentration of 0.1 M, adjusting the number of ions to ensure an overall net neutral charge in the system. For the initial configuration, we set the protein in the center of a dodecagonal box with periodic boundaries, with a minimum distance of 1 nm between the protein and the box edges. We use the OPLS forcefield for proteins and ions, and the SPC model for water. Interaction cutoffs of 1 nm are used for electrostatics. After energy minimization, systems are equilibrated by running 100 ps in the NVT ensemble, followed by 100 ps in the NPT ensemble; T = 300 K, P = 1 bar; velocity-rescale thermostat and Parrinello-Rahman barostat are used. Leapfrog integration (2 fs timestep) and Verlet neighbor lists were used. Statistics were gathered over 100 ns. For proteins nirD and cutinase, we used the AlphaFold-predicted structures as initial configurations [75] since the PDB structures were missing several atoms.

**MCNP Simulations.**

The system was constructed similarly to our previously published simulations.[76] The metallic core of AuNP was modeled as a perfect icosahedron of a maximum diameter of 4.4 nm, comprised of 2057 Au atoms, whose coordinates were obtained with the icosahedralBuilder script available in the OpenMD package.[77] Next, the *N,N,N*-trimethyl(11-mercaptoundecyl)ammonium chloride (TMA) and 11-mercaptoundecanoic acid (MUA) ligands were attached to the surface of AuNP with thiol groups so that the resulting distribution was approximately homogenous − that is, neither ligands tend to form 'clusters' on the surface of NP. TMA ligands are positively charged (net charge +1) and MUA ligands are deprotonated (net charge -1).

Depending on the total charge of the whole nanoparticle, Cl- or Na+ ions were added as counterions to neutralize the system.

The positions of sulfur atoms fulfilling the homogeneity criterion were obtained by in-house script computing the solution to the Thomson problem, where the objective is to find a minimum energy distribution of N mutually repelling 'point charges' on a sphere (because of spherical symmetry, such solution tends to be uniform). The loss function applied here is $E = \sum r_{ij} - 1$, where $r_{ij}$ denotes the distance between points $i$ and $j$, the summation runs over all unique pairs, and each point is constrained to the sphere's surface.

Here, we want to distribute $m + n$ points belonging to two classes (MUA and TMA, with sizes $m$ and $n$, respectively), so neither class 'clumps' on the sphere. This is computed in two steps: first, the positions of MUA sulfur atoms alone are optimized with gradient descent. Secondly, with MUA sulfur atoms' positions kept fixed, the TMA positions are optimized (the loss function at this stage encompasses contributions from all points, but only TMA are allowed to move). Finally, the resulting distribution is radially moved such that the sulfur atoms touch the surface of the Au core. The value of $m + n$ was taken as 200, which corresponds to number-averaged surface density ~5.2/nm$^2$. Attempts to initialize the structure with higher ligand density led to the generation of 'short-contacts' in the coordinates (i.e., the shortest distances between monomers < 1 Å), which might lead to the failure of subsequent simulation.

All simulations were performed with the Gromacs 2022.4 package.[74] The NP was placed in a dodecahedral box with the face-AuNP margin set to 2.5 nm. All models were solvated with the TIP3P water model, and chloride ions were added to maintain charge neutrality. The GAFF force field was used in all simulations.[78] Parameters for gold atoms were taken from the work of Heinz and co-workers,[79] whereas for the MUA and TMA ligands, they were generated with the antechamber feature from AmberTools22.[80] Positions of all gold atoms were constrained during simulations. Each system was subject to energy minimization (with the steepest descent algorithm), NVT (100 ps long) and NPT (200 ps) equilibration phases, and the final production phase (1 ns), from which the average interaction energies and other parameters were collected.

**MCNP surface analysis**

We modified the procedure from the Section **"Protein surface analysis"**. "Reduce" and PDB2PQR are not needed since the MD simulations allow the direct creation of .pqr files with hydrogen atoms. We use MSMS to generate the solvent-accessible surface and use APBS to evaluate the surface potential. We analyze surface curvature and patchiness for NPs like we do for proteins.

**Statistical Analysis**

Pearson's correlation coefficients are reported in **Fig. 2C**; sample sizes are equivalent to protein lengths, which typically range from 100 to 300 amino acids

# Results

### General features of amino acid distribution on protein surfaces.

Amino acid distributions within proteins and on their surfaces have been extensively studied,[18,81] so we begin by providing a detailed update of these statistics. Thus, we first summarize the fractions of amino acids on protein surfaces (**Fig. 2A**). We grouped the 20 amino acids into five classes: 5 charge-neutral polar (we denote the amino acids by their single letter odecode: S, T, N, Q, Y), 3 positively charged (R, K, H; H is pH-responsive with $pK_a = 6$), 2 negatively charged (D, E), 5 weakly hydrophobic (G, A, P, M, W), and 5 strongly hydrophobic (V, C, L, I, F). In general, we find that these five classes are approximately evenly distributed amongst protein surfaces. The least abundant class is of the strongly hydrophobic amino acids. However, it also has a clear negative skew: some membrane proteins have many more of these amino acids than average on the surface.[29] In comparison, charged groups exhibit much broader distributions, showing much more variability between proteins. Compared to the statistics of whole protein sequences, protein surfaces are enriched in charged amino acids, depleted in hydrophobic residues (strong hydrophobic more depleted than weak), and quite surprisingly, show no difference for polar residues.[82]

### Surface potential is the most variable property on protein surfaces.

The mean values of hydrophobicity and the magnitude of the electric potential on the surface (Mag. Potential) are two complementary coarse-grained indicators of surface chemistry (**Fig. 2B**). Most proteins maintain a mean score of ~ -1.3 on the Kyte-Doolittle hydrophobicity scale (which ranges from −4.5, most hydrophilic, to 4.5, most hydrophobic). Membrane proteins (GO:0016021) are an exception – they tend to be more hydrophobic, though they rarely have a mean score above 1. The most significant variation in protein surfaces is seen in surface potential: while most proteins (72%) have net-negative surfaces, a significant fraction is net-positive (as defined by the mean of the electric potential). This variation is linked to function: *e.g.*, nucleic-acid-binding proteins (GO:0003676) tend to have more positively charged surfaces since nucleic acids possess a negatively charged backbone.[83] Despite this variation in mean surface potential, the magnitude of the surface potential for water-soluble proteins is roughly bound between 5 kT/e and 10 kT/e. We find that protein surface properties vary depending on subcellular location and function, and according to taxa (**SI Fig. 2-3**). To give a few examples, extracellular proteins and animal/fungi proteins tend to have more polar residues on the surface (**Fig. 2B**, **SI Fig. 3**), animals tend to have more negatively charged proteins (**SI Fig. 3**), while bacterial proteins are more hydrophilic. However, the variation between these groups is ultimately small compared to the variation within each group.

### Curvature correlates with charge and hydrophobicity.

One expects to find that certain surface properties are intrinsically correlated. For example, hydrophobicity and charge should be anti-correlated, as shown in **Fig. 2C**, which displays the distribution of correlation coefficients measured between the surface properties of each protein. A null example of practically no correlation is the distribution for mean curvature and surface potential. Among the possible non-intrinsic correlations, we show two that are most significant: mean curvature vs., respectively, hydrophobicity and magnitude of electrostatic potential (Mag. Potential). Concave regions characterized by negative mean curvature are relatively more hydrophobic, and convex regions of high positive mean curvature tend to be more charged. Such correlations between hydropathy, charge, and curvature are expected: concave regions are less solvent exposed, and if the hydrophobic core interacts with the solvent, this will likely occur in a concave region; on the other hand, charged regions favor interactions with solvent, which can be maximized at a convex surface.

### Surface properties vary over short length scales.

The function of proteins relies on the spatial patterns of their surface properties. To probe the length scales at which the properties vary, we calculated the autocorrelation as a function of the geodesic distance (the shortest distance between two points on a surface). We find that curvature varies over the shortest length scales (**Fig. 2D**): Gaussian curvature has a length scale of ~ 2 Å, and mean curvature is correlated over slightly longer lengths of about 5 Å (compared to previously estimated curvature length scale of ~ 7 Å [84]). Hydrophobicity typically varies over a distance of at least 10 Å, roughly the length scale of a single amino acid. This suggests that, usually, hydrophobic residues do not form large clusters and are well distributed over the surface. We find that the surface potential autocorrelation decays as $1/r$ (**Fig. 2D**), as the potential of an isolated point charge, implying that, on average, like charges do not cluster on the protein surface, probably due to repulsion.

### Patches of surface properties exhibit diverse sizes, shapes, and 3D arrangement.

While the correlation function reasonably estimates the average length scale, it misses important geometric features of the surface patches, such as their aspect ratio. Therefore, to study surface patches in more depth, we divide the surface into connected regions of high, low, and medium values of the various surface properties (**Figure 1**). We quantify the shape of these clusters using a topological measure $D$ termed "discosity," where $D \to 0$ is for elongated clusters and $D \to 1$ is for disc-like clusters (see Methods). The discosity reflects the cluster's area-to-perimeter ratio. This clustering approach gives us a clear view of how patches vary in their curvature (**Fig 2E**). Regions of high Gaussian curvature (i.e., highly convex or concave) tend to be small and disc-like, while areas of medium Gaussian curvature (approximately flat) are small and elongated. Clusters of low Gaussian curvature (flat areas and saddle points) tend to dominate, forming larger extended clusters. Clusters of mean curvature provide further geometric insight. The regions with medium-high Gaussian curvature are split into those with positive (convex) and negative (concave) mean curvature. Convex regions are mainly disc-like patches, while concave regions are typically smaller and more elongated. Put another way, protein surfaces have knob-like protrusions, but the concave regions tend to be more like troughs than dimples. Overall, this paints a picture of protein surfaces as manifolds with rapidly changing curvature, composed of small patches of knobs and troughs surrounded by relatively larger flat or saddle-like surfaces.

As for chemical features, the patch geometry is less clearly distinguished. Protein surfaces are mainly hydrophilic, and thus hydrophilic clusters are typically large (mean 654 Å$^2$) and tend to have high discosity. There is little difference in the size of medium- and high-hydrophobicity clusters (113 Å$^2$ and 100 Å$^2$, respectively), but high-hydrophobicity clusters tend to be more disc-like. For surface potential, we labeled the surface potential as medium between +/−5 kT/e, where 5 kT/e corresponds to the average magnitude of the potential (**SI Fig. 1**). Little difference is found in shape or size between positive (49 Å$^2$) and negative (85 Å$^2$) regions, which are dwarfed by the region with medium surface potential (411 Å$^2$). This size difference is expected owing to the energetic cost of placing like-charges together, which may hinder protein folding. Moreover, capping the size of hydrophobic, positively charged, or negatively charged patches is a practical design principle since, otherwise, they are likely to make proteins aggregate. Further insights may come from looking at patch-patch correlations, where one might distinguish something like hydrophobic/electrostatic channels.[85]

The overall shape of proteins is also important for comparison with nanoparticles. In brief, proteins can take on many shapes (*e.g.*, spheres, doughnuts, bananas) and sizes,[86] but are generally closest to ellipsoids,[87–89] more prolate than oblate,[90,91] with an effective diameter of ~ 2−6 nm.

### Mixed-charge nanoparticles (MCNPs)

With these broad statistics of protein surfaces in mind, we compare proteins to mixed-charge nanoparticles (MCNPs). The MCNPs are gold nanoparticles functionalized with controllably-varied mixtures of n-alkane thiolate ligands terminated in either positive (trimethylammonium, TMA) or negative (carboxylic acid, MUA) head groups. The alkane linkers typically contain 11 or 10 carbon atoms for TMA and MUA ligands, respectively. Of note, MUA residues are pH-responsive - they become protonated at pH ~<6 [92]. For details on synthesis, see Borkowska et al.[55] At first glance, the MCNPs are similar to proteins in size (<10 nm) and shape (approximately spherical), and both are covered with small patches of charged groups. To obtain a more quantitative, detailed view of MCNPs, we performed MD simulations and analyzed geometry, dynamics, and surface charge/hydrophobicity.

### Comparing MCNPs to proteins: geometry & dynamics

While proteins and MCNPs tend to be approximately spherical, MCNPs appear to have a distinctive hairy-ball topology. Protein surfaces are rough and tend to have large cavities for binding; in comparison, the MCNP surfaces have deep voids that form an interconnected network (**Fig. 3A**). At smaller length scales, the surface roughness of MCNPs is very similar to proteins as they are composed of surface patches of similar curvature (compare **Figs. 2E, 3B**). The MCNP surface topology is partly a reflection of the molecular construction – multiple ligands attached to a spherical gold NP. Although exposure of the hydrophobic linkers is unfavorable in water, the linkers are not flexible enough to collapse and thereby reduce the exposed surface area. Importantly, there are a couple of possible ways to change the strength of interactions between linker chains. First, changing the NP core's size/local curvature will change the relative angles between linker chains – at higher curvatures, the spaces between ligands are bigger and vice versa. The second option is changing the chemical composition of the ligand, i.e., adding ethylene glycol moieties in the chain to make it more hydrophilic and flexible or changing the anchoring group to change the surface density of ligands.

The dynamics of MCNPs is quite similar to proteins. Both MCNPs and the three proteins studied here exhibit root mean square fluctuations (RMSF) in atomic positions of about 0.3 nm (**Fig. 3C**). However, proteins can undergo large conformational changes over long times, which is not possible in NPs due to their constrained molecular topology; the largest possible movements are when neighboring ligands change their orientation away from the NP surface, of which the largest observed shift is 30 degrees (**SI Fig. 4**).

We observe slight differences in MCNPs depending on the ligand composition. To estimate ligand packing, we calculated the distance between all pairs of atoms in neighboring ligands, from which we obtained an average distance (the head groups contain more atoms and thus have a higher weight in the average). This average distance indicates that MUA ligands are bunched closer together and more ordered than TMA ligands (**SI Fig. 5**), probably due to both their ability to form hydrogen bonds between each other and due to the decreased steric repulsion compared to TMA ligands, which have bulkier head-groups. This results in narrower, strongly connected voids (**Fig. 3A**; larger clusters of negative Gaussian curvature, **Fig. 3B**) in MCNPs with more MUA ligands, and also greater flexibility in MCNPs with more TMA ligands. Changing the strength and nature of ligand-ligand thus offers a way to modulate the geometry and dynamics of MCNP surfaces.

**Comparing MCNPs to proteins: surface charge and hydrophobicity**

While MCNPs and proteins both have a mix of positively and negatively charged moieties on the surface, the electrostatic environments differ quite significantly. The ligands used to create MCNPs all have charged head groups at physiological pH, whereas charged amino acids typically cover less than half of protein surfaces (**Fig. 2A**). This results in the charge-imbalanced NPs having considerably higher surface potential than proteins, although it is possible to experimentally tune the NP potential to be around zero for MCNPs with 50:50 ratio of ligands. (**Fig. 3D**). Despite significant differences in net surface potential, the NP surface potential appears to be as heterogeneous as typical protein surfaces (**SI Fig. 6**). In contrast, there are much fewer opportunities for hydrogen bonding on the surfaces of the TMA/MUA MCNPs we examined (**Fig. 3D**, **SI Fig. 7**) since ammonium salts do not form hydrogen bonds. Extrapolating from this, by changing the chemistry of the ligand shell, it would be possible to fine-tune its properties (e.g., by using tri-carboxylic acids instead of MUAs, we could significantly increase the possibility of forming hydrogen bonds, whereas by using neutral ligands we could decrease the net charge density).

One similarity between MCNPs and proteins is the presence of hydrophobic cavities. In MCNPs examined, these cavities have high electric potential due to the high charge density of the ligand heads, but there are no isolated charges or opportunities for water to interact other than through van der Waals interactions. The formation of hydrophobic cavities is a direct result of the linker rigidity and the spherical geometry of the MCNP core. The hydrophobic linkers are attracted to each other by van der Waals forces, so they bunch up. Since the volume of a spherical shell increases as its radius squared, but the linker cross-section remains constant, voids are bound to appear and expose the hydrophobic linkers to solvent. Thus, this geometric effect can be exploited to control the size of hydrophobic cavities by changing the size and shape of the NP core or the structure and/or composition of the linkers.

## Discussion

### Function and specificity in proteins and MCNPs

Protein surfaces have evolved to function efficiently within the elaborate biomolecular circuitry of the living cell. Within this noisy milieu, proteins need to adapt their interactions to have optimal values of *specificity* and *affinity* so that they bind to correct targets (function) while maintaining favorable interactions with the appropriate solvent (solubility) and minimizing binding to incorrect targets (cross-talk, aggregation, side-reactions).[23] Properties such as solubility derive from broad statistical distributions of surface "patches" of positive/negative charges and/or hydrophobic/hydrophilic groups. Large hydrophobic or charged patches can lead to nonspecific, high-affinity binding or undesired aggregation. Likewise, guest molecules are less likely to be recognized as foreign by the immune system if their surfaces are chemically and geometrically similar to host molecules. By contrast, specificity and activity rely on the delicate arrangement of a small number of amino acids (especially in enzymatic catalytic sites). Thus, for most proteins, function is determined by a mix of *statistical* and *fine-tuned* features.

The MCNPs we focused on in this work also mostly exhibit statistical heterogeneity, in terms of +/- patchiness, as well as molecular-scale topology with head-groups at the exterior surface and linkers at the interior surface. The fine-tuned features are less controllable on MCNPs, though it should be noted that with the extension of the ligand repertoire, some rough mimics of catalytic sites have been demonstrated, as in the so-called "nanozymes".[93–97] For instance, we have previously showed[76,98,99] that longer charged ligands can serve as "gating" units controlling access to shorter ligands terminated in catalytically active groups – in effect, only certain small molecules can reach the catalytic sites and in specific orientations, ensuring regioselective reaction outcomes. Similar schemes can be further diversified to different types of catalytic and gating units and, conceivably, to lateral phase separation[100–102] such that qualitatively different catalytic microenvironments on the same NP could be realized (which would be unlike in most proteins that feature one active site).

Another worthwhile aspect regards the evolution of some proteins to be quite promiscuous in their interactions, as one protein must be able to regulate many other proteins, thus minimizing the machinery needed for proteostasis.[103] A prominent example of such generalist proteins are chaperones that recognize unfolded or misfolded proteins by their typical large hydrophobic patches that are rare in native structures (**Fig. 2E**). Chaperon proteins typically have an extended hydrophobic region placed in a cavity, surrounded by charged patches, that is used to recognize such incorrectly folded structures.[104] In this respect, we find that the MCNPs exhibit somewhat similar surface patterns and thus may be promising candidates for *artificial chaperones*. Such NPs would potentially facilitate the storage and transport of proteins for biomedical and research applications, especially since protein shelf life is typically cut short by an inability to refold *in vitro* and at high concentrations.

Having narrated the similarities, it is essential also to emphasize the key differences between proteins and MCNPs (or most other synthetic NPs):(i) the existence of a very rigid, homogeneous core (metallic or non-metallic, e.g., $SiO_2$ or $TiO_2$), typically generating strong vdW interactions (~10 kT for metals)[105] and (ii) much higher stability (also thermal [106,107]) of the ligand shell compared to the conformation-dependent surface properties of proteins.

Proteins consist of a single linear amino acid chain that folds into a stable tertiary structure. Although polymeric NPs can mimic this topology,[108–113] many NPs have a ligand-core or dendritic structure. This branched topology should result in different surface dynamics. Protein dynamics are mostly constrained to vibrations, local folding/unfolding, and a few large-scale coordinated movements, such as in allosteric transitions.[7,114–117] In contrast, NPs surfaces may be fluid-like to glass-like, depending on the degree of packing of ligands on the NP shell and the strength of the ligand-ligand and ligand-solvent interactions.[99]

Proteins are encoded in long genes (typically thousands of DNA bases) and are produced by transcribing the gene into a messenger RNA, which is then translated at the ribosome,[118,119], achieving a phenomenally low error rate of 1 per 1,000-10,000 amino acids.[39,120] The resulting polypeptide chain must then fold into a functional protein (except for intrinsically disordered proteins that remain unfolded). At this stage, some fraction of proteins misfold into long-lived, partially-folded states, but high purity can be achieved by proteolytic degradation that targets misfolded forms. Thus, the end product is typically a highly monodisperse sample of proteins: almost identical molecules with almost identical structures. The main drawback of such a highly optimized process is that it is exceptionally challenging to include amino acids beyond the 20 encoded by most genomes,[121,122] and that the error correction mechanisms are costly. By contrast, synthetic NPs offer – at least in principle – more design flexibility in terms of chemical groups present on the surface, albeit it should be stressed that the vast majority of NPs described in the literature have been using monocomponent or just a few-component shells.

The similarities and differences between proteins and NPs are summarized in **Table 1**.

| Property | Similarities | Differences |
|---|---|---|
| Surface chemistry | Combinatorial potential for variability (amino acids vs. ligands). | NPs have a greater range of ligands to choose from. Proteins can use ~ 20 amino acids easily but are mostly limited to this set; proteins are translated from genes and |

|  |  | can also be chemically modified post-translationally (e.g., glycosylation). |
|  | Control of statistical distribution of surface properties. | Proteins can specifically control the spatial distribution of amino acids. Control of spatial distribution on NPs is limited to particular cases (e.g., a certain ligand having a much higher affinity to only one crystal facet or edge[123]). |
| Surface dynamics | The surface will fluctuate, but ligands / amino acids are mostly fixed in place. | Long-range conformational changes allow proteins to toggle between structural states. |
|  |  | NPs can tune surface dynamics by controlling the structure and packing of ligands. |
| Shape and Size | A variety of geometries are possible, but most popular are globular-like shapes. | NPs offer a large range of sizes and precise control of aspect ratios and shapes. Proteins are much more monodisperse and fine-grained control - their shapes vary widely in elongation, convexity, etc.; however, they are size-limited as protein folding becomes more challenging at large sizes. |

**Table 1.** Similarities and differences between proteins and NPs.

### Design blueprints for protein-mimetic NPs

One of our main motivations in this work has been to use the knowledge gained from the survey of protein surfaces for the design of more protein-like NPs. While methods for nanoparticle synthesis certainly cannot replicate the intricacies of protein shape, they are sufficient to reproduce the protein's 2-10 nm sizes and rough contours, be it spherical or ellipsoidal[124] or even bowl-shaped.[125] Non-convex shapes are particularly attractive for mimicking the function of chaperone proteins,[126] which tend to have solvent-exposed hydrophobic patches within cavities.[127–129]

NP surface chemistry significantly affects intermolecular interactions, which in turn determine the likely extra- and sub-cellular environments they will accumulate in, whether through the formation of protein coronas,[130,131] propensity towards different chemical environments, or differences in membrane crossing rates.[132] In terms of the dominant hydrophilicity of proteins' outer surfaces and the hydrophobicity of concave isolated patches, this can be largely translated into NPs by using only ligands with hydrophobic chains and hydrophilic end groups, with hydrophilic/hydrophobic surface ratio tunable by choice of ligand length and head-group size. The size and shape of the hydrophobic cavities can also be tuned by the shape and size of the NP core, the length and chemistry of the linkers, and the choice of head-groups: surfaces with higher curvature will lead to wider voids, longer linkers will increase void volume, and attraction between head-groups will lead to wider voids between clusters of ligands. Other possible means of control include using linkers incorporating hydrophilic (e.g., ethylene glycol) units to decrease the hydrophobicity of the voids.

In principle, tuning the bulkiness and lengths of mixed on-nanoparticle ligands may lead to microphase separation[101,133,134] and the formation of larger clusters of charges. One of the most precise measurements, using MALDI-IM-MS by Harkness et al.[135], confirms that the phase separation is more prominent when mixed ligands are of different lengths, and their mutual interactions are limited.

Finally, regarding charge density, we find that proteins tend to have ~4 hydrogen-bonding sites (HBS) per nm$^2$ (**Fig. 3B**, **SI Fig. 7**), and the ratio of charged amino acid HBS to polar HBS is about 5:3. Thus, in protein-mimicking NPs, the head-groups with single charge centers may be sufficient in many cases, as protein surfaces are not very strongly charged (**Fig. 3B**). Additionally, the charge density on NPs can be easily tuned by adding the "dilution ligand".[57]

## Conclusion

We have provided a detailed, statistical view of the geometric and chemical features of protein surfaces. By directly comparing proteins with mixed-charge nanoparticles (MCNPs), we could catalog the basic similarities and differences and make recommendations for how to design protein-mimetic NPs. Proteins and NPs are generally similar in their compact nano-scale geometry, but they differ significantly in the underlying topology: proteins are intricately folded amino acid chains, whereas the NPs studied here are hairy balls, solid cores covered with linear or branched ligands. Dynamics at short time scales are, therefore, quite similar, although for the particular NP topology studied, it is unlikely that NPs will mimic the slow, long-ranged conformational changes with local folding\unfolding involved in protein allostery[7,114–116]. We observed significant differences in terms of surface chemistry, mainly net charge, but this is a feature that is easily tunable through altering the composition of surface ligands.

Despite these differences, NP and protein surfaces consistently feature islands of charge, surrounded by meandering networks of hydrophobic cavities. In this work, we focused on proteins and compared them to a single family of MCNPs as a case study. Future work should systematically characterize the surfaces of NPs designed for use in vivo and investigate the effect of post-translational modifications (e.g., glycosylation) on proteins; to facilitate this, we release the code used in this work at github.com/jomimc/masif_molecule. Such a systematic survey could potentially uncover statistical regularities between NPs that are or are not suitable for in vivo use and restrict the design space of NPs for such applications. We envision that in the fullness of time, such recommendations and their future extensions can help the design of NPs with tailored solubilities, toxicities, and specificities of the binding sites, and with applications in drug delivery, NP therapeutics, industrial-scale production of stable proteins, and, by extension, catalysis design.

**Data availability statement**

Code used to extract surfaces and surface features is available at https://github.com/jomimc/masif_molecule.

**Acknowledgments**

This work was supported by the Institute for Basic Science, Project Code IBS-R020-D1.

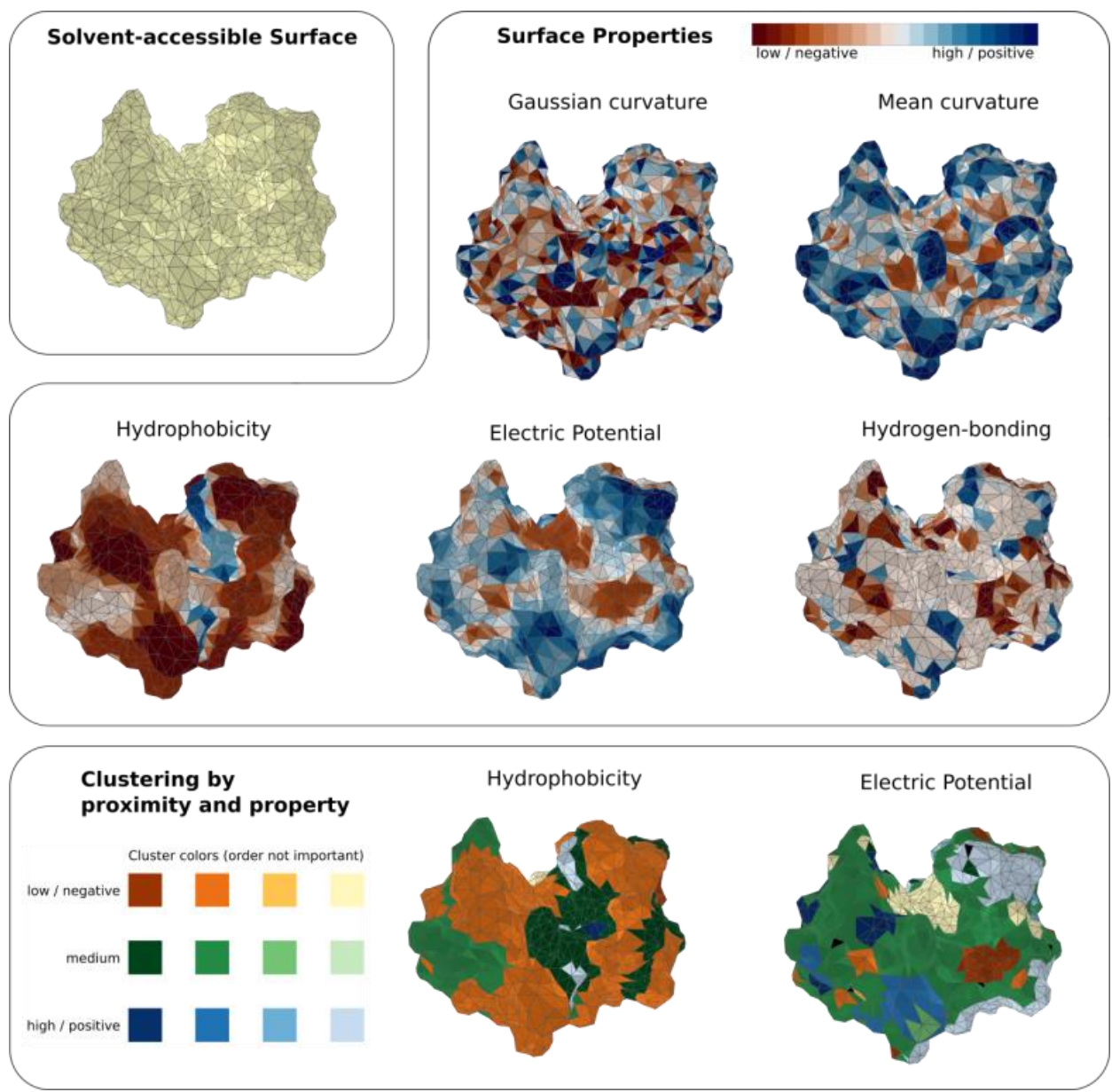

**Figure 1**

The solvent-accessible surface is represented by a triangulation, whereby each vertex is labeled with two geometric and three chemical properties: Gaussian and mean curvatures; hydrophobicity, electric potential, and hydrogen bonding potential. We cluster the surface according to regions with similar properties (**SI Fig. 1**) that are connected. We show hen lysozyme (PDB, 194L) as an example of a protein with a strongly hydrophilic, positively charged surface.

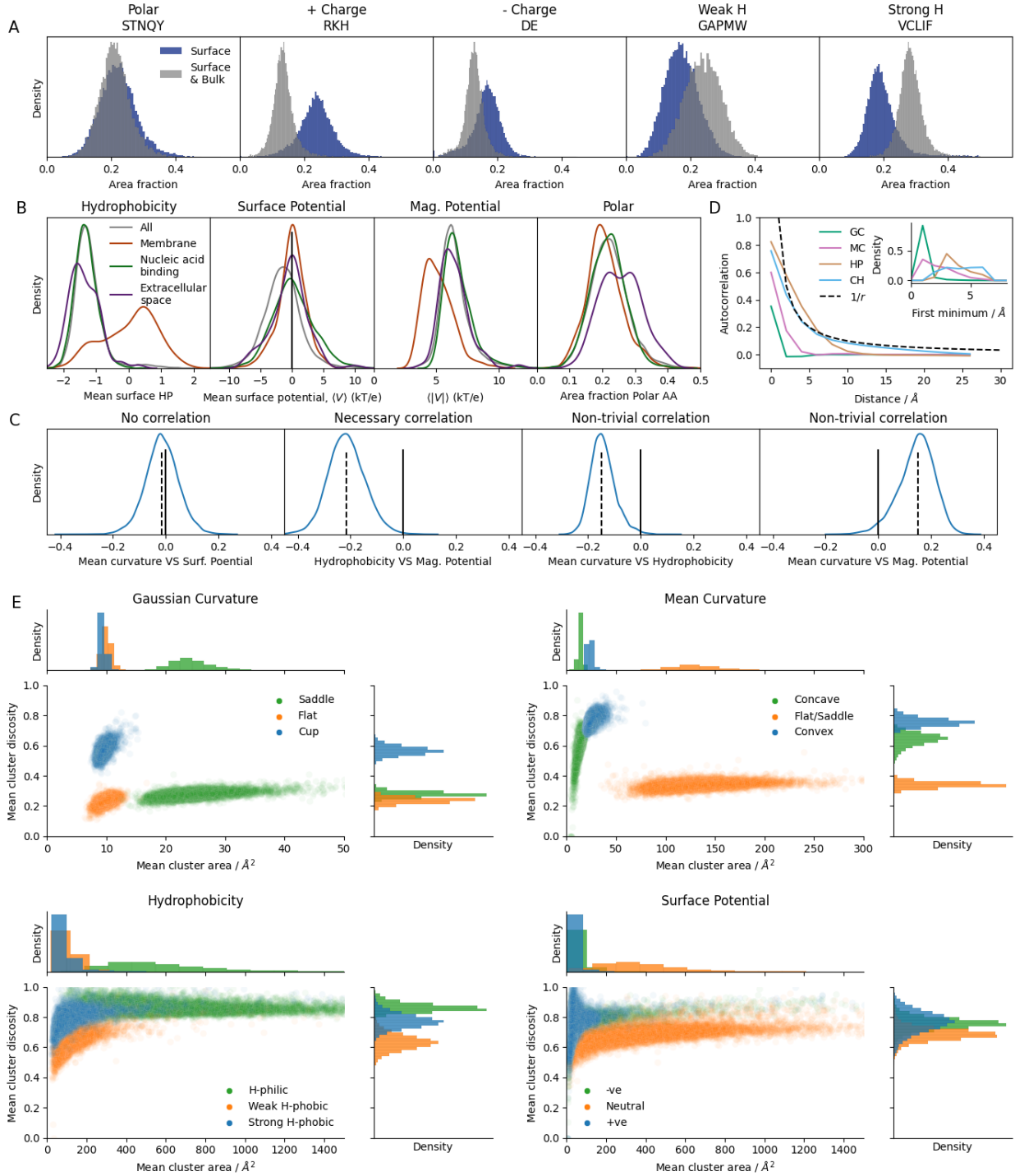

**Figure 2**

**A**: Distributions of surface area fraction for different amino acid groups (left to right): polar (serine, threonine, asparagine, glutamine, tyrosine), positively charged (arginine, lysine, histidine), negatively charged (aspartic acid, glutamic acid), weakly hydrophobic (glycine, alanine, proline, tryptophan, methionine), and strongly hydrophobic (valine, cysteine, leucine, isoleucine, phenylalanine). Distributions are shown for the surface of the proteins and for the entire protein volume. **B**: Distributions (kernel density estimates) of average surface properties: Kyte-Doolittle hydrophobicity scale (HP); surface potential (CH); magnitude of surface potential (Mag. Potential); surface area fraction of polar amino acids (Polar). Distributions are shown for: all proteins, membrane-associated proteins (GO:0016021), nucleic acid-binding proteins (GO:0003676), and extracellular proteins (GO:0005615). A solid black line is a guide to the eye. **C**: Distribution of correlation coefficients between surface properties (one coefficient is calculated per protein surface). A black solid denotes the zero on the x-axis; the black dashed line shows the mean of the distribution. **D:** Autocorrelation function of properties on the surface as a function of the geodesic distance on the surface, averaged over all proteins. Inset: Distributions of the first minima in the autocorrelation function; Gaussian curvature (GC), mean curvature (MC). **E:** Distribution of mean size and mean discosity for clusters of similar properties on a protein surface: Gaussian curvature, mean curvature, hydrophobicity, and surface potential.

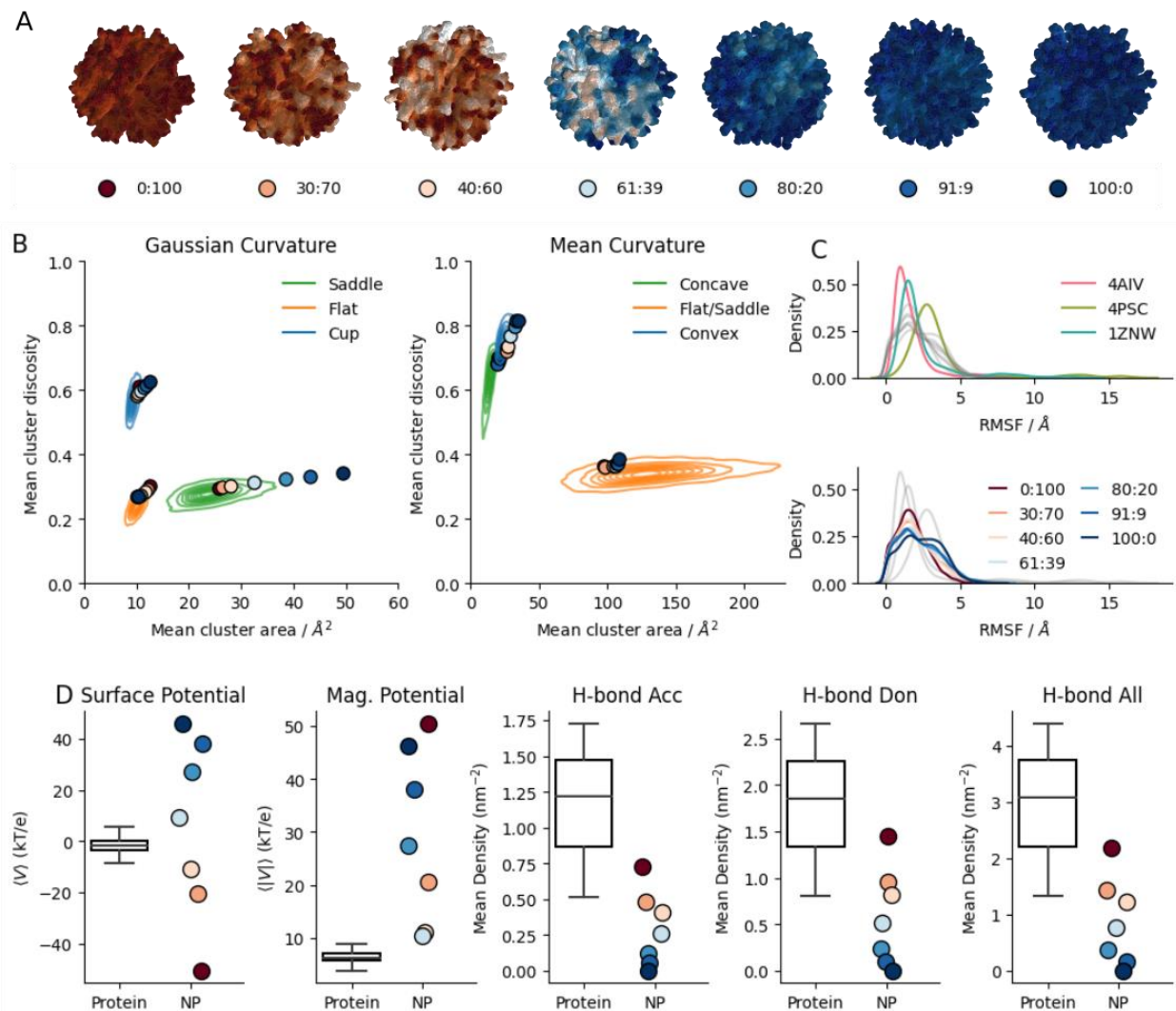

**Figure 3**

**A:** Snapshots of solvent-accessible surfaces colored according to the surface potential for MCNPs of different TMA:MUA ligand compositions.

**B:** Distribution of mean size and mean discosity for clusters of similar properties on a protein surface: Gaussian curvature, mean curvature. Contour lines indicate distributions for proteins; values for MCNPs are shown as points, colored according to TMA:MUA ligand ratio.

**C:** Distribution of root mean square fluctuations (RMSF) per atom for proteins and MCNPs. (Top) Three proteins (nirD, *mycobacterium tuberculosis*, PDB ID 4AIV; cutinase, *hypocrea jecorina*, PDB ID 4PSC; guanylate kinase, *mycobacterium tuberculosis*, PDB ID 1ZNW) are indicated by color, and MCNPs are shown in grey. (Bottom) MCNPs of different ligand compositions are colored according to TMA:MUA ligand ratio, and proteins are shown in grey.

**D:** Box plots show distributions of mean surface properties for proteins, and mean values for each MCNP are shown as circles colored by ligand composition.

## References

(1) El-Gebali, S. The Pfam Protein Families Database in 2019. *Nucleic Acids Res* **2019**, *47*, 427–432.
(2) Knudsen, M.; Wiuf, C. The CATH Database. *Hum Genomics* **2010**, *4*, 207.
(3) UniProt Consortium. UniProt: The Universal Protein Knowledgebase in 2021. *Nucleic Acids Res.* **2021**, *49* (D1), D480–D489. https://doi.org/10.1093/nar/gkaa1100.
(4) Berman, H. M. The Protein Data Bank. *Nucleic Acids Res* **2000**, *28*, 235–242.
(5) Andreeva, A.; Kulesha, E.; Gough, J.; Murzin, A. G. The SCOP Database in 2020: Expanded Classification of Representative Family and Superfamily Domains of Known Protein Structures. *Nucleic Acids Res* **2020**, *48*, 376–382.
(6) Kessel, A.; Ben-Tal, N. *Introduction to Proteins: Structure, Function, and Motion*; CRC Press, 2010.
(7) McBride, J. M.; Tlusty, T. The Physical Logic of Protein Machines. *J. Stat. Mech. Theory Exp.* **2024**, *2024* (2), 024001. https://doi.org/10.1088/1742-5468/ad1be7.
(8) Ashburner, M. Gene Ontology: Tool for the Unification of Biology. *Nat Genet* **2000**, *25*, 25–29.
(9) Jumper, J.; Evans, R.; Pritzel, A.; Green, T.; Figurnov, M.; Ronneberger, O.; Tunyasuvunakool, K.; Bates, R.; Žídek, A.; Potapenko, A.; Bridgland, A.; Meyer, C.; Kohl, S. A. A.; Ballard, A. J.; Cowie, A.; Romera-Paredes, B.; Nikolov, S.; Jain, R.; Adler, J.; Back, T.; Petersen, S.; Reiman, D.; Clancy, E.; Zielinski, M.; Steinegger, M.; Pacholska, M.; Berghammer, T.; Bodenstein, S.; Silver, D.; Vinyals, O.; Senior, A. W.; Kavukcuoglu, K.; Kohli, P.; Hassabis, D. Highly Accurate Protein Structure Prediction with AlphaFold. *Nature* **2021**, *596* (7873), 583–589. https://doi.org/10.1038/s41586-021-03819-2.
(10) Baek, M.; DiMaio, F.; Anishchenko, I.; Dauparas, J.; Ovchinnikov, S.; Lee, G. R.; Wang, J.; Cong, Q.; Kinch, L. N.; Schaeffer, R. D.; Millán, C.; Park, H.; Adams, C.; Glassman, C. R.; DeGiovanni, A.; Pereira, J. H.; Rodrigues, A. V.; Van Dijk, A. A.; Ebrecht, A. C.; Opperman, D. J.; Sagmeister, T.; Buhlheller, C.; Pavkov-Keller, T.; Rathinaswamy, M. K.; Dalwadi, U.; Yip, C. K.; Burke, J. E.; Garcia, K. C.; Grishin, N. V.; Adams, P. D.; Read, R. J.; Baker, D. Accurate Prediction of Protein Structures and Interactions Using a Three-Track Neural Network. *Science* **2021**, *373* (6557), 871–876. https://doi.org/10.1126/science.abj8754.
(11) McBride, J. M.; Polev, K.; Abdirasulov, A.; Reinharz, V.; Grzybowski, B. A.; Tlusty, T. AlphaFold2 Can Predict Single-Mutation Effects. *Phys. Rev. Lett.* **2023**, *131* (21), 218401. https://doi.org/10.1103/PhysRevLett.131.218401.
(12) Mcbride, J. M.; Tlusty, T. AI-Predicted Protein Deformation Encodes Energy Landscape Perturbation. *ArXiv Prepr. ArXiv231118222* **2023**.
(13) Huang, P.-S.; Boyken, S. E.; Baker, D. The Coming of Age of de Novo Protein Design. *Nature* **2016**, *537*, 320–327.
(14) Fersht, A. *Structure and Mechanism in Protein Science: A Guide to Enzyme Catalysis and Protein Folding*; Macmillan, 1999.
(15) Nakamura, H. Roles of Electrostatic Interaction in Proteins. *Q Rev Biophys* **1996**, *29*, 1–90.
(16) Liang, J.; Edelsbrunner, H.; Woodward, C. Anatomy of Protein Pockets and Cavities: Measurement of Binding Site Geometry and Implications for Ligand Design. *Protein Sci* **1998**, *7*, 1884–1897.
(17) Dundas, J. CASTp: Computed Atlas of Surface Topography of Proteins with Structural and Topographical Mapping of Functionally Annotated Residues. *Nucleic Acids Res* **2006**, *34*, 116–118.
(18) Yan, C.; Wu, F.; Jernigan, R. L.; Dobbs, D.; Honavar, V. Characterization of Protein–Protein Interfaces. *Protein J* **2008**, *27*, 59–70.
(19) Scott, D. E.; Bayly, A. R.; Abell, C.; Skidmore, J. Small Molecules, Big Targets: Drug Discovery Faces the Protein–Protein Interaction Challenge. *Nat Rev Drug Discov* **2016**, *15*, 533–550.
(20) Hadarovich, A. Structural motifs in protein cores and at protein–protein interfaces are different. *Protein Sci* **2021**, *30*, 381–390.

(21) Hermann, J. C. Structure-Based Activity Prediction for an Enzyme of Unknown Function. *Nature* **2007**, *448*, 775–779.
(22) Gainza, P. Deciphering Interaction Fingerprints from Protein Molecular Surfaces Using Geometric Deep Learning. *Nat Methods* **2020**, *17*, 184–192.
(23) McBride, J. M.; Eckmann, J.-P.; Tlusty, T. General Theory of Specific Binding: Insights from a Genetic-Mechano-Chemical Protein Model. *Mol. Biol. Evol.* **2022**, *39* (11), msac217. https://doi.org/10.1093/molbev/msac217.
(24) Kim, Y. C.; Tang, C.; Clore, G. M.; Hummer, G. Replica Exchange Simulations of Transient Encounter Complexes in Protein–Protein Association. *Proc Natl Acad Sci* **2008**, *105*, 12855–12860.
(25) Warfield, L.; Tuttle, L. M.; Pacheco, D.; Klevit, R. E.; Hahn, S. A Sequence-Specific Transcription Activator Motif and Powerful Synthetic Variants That Bind Mediator Using a Fuzzy Protein Interface. *Proc Natl Acad Sci* **2014**, *111*, 3506–3513.
(26) Sharma, R.; Raduly, Z.; Miskei, M.; Fuxreiter, M. Fuzzy Complexes: Specific Binding without Complete Folding. *FEBS Lett* **2015**, *589*, 2533–2542.
(27) McCarty, J.; Delaney, K. T.; Danielsen, S. P. O.; Fredrickson, G. H.; Shea, J.-E. Complete Phase Diagram for Liquid–Liquid Phase Separation of Intrinsically Disordered Proteins. *J Phys Chem Lett* **2019**, *10*, 1644–1652.
(28) Espinosa, J. R. Liquid Network Connectivity Regulates the Stability and Composition of Biomolecular Condensates with Many Components. *Proc Natl Acad Sci* **2020**, *117*, 13238–13247.
(29) Ulmschneider, M. B.; Sansom, M. S. P. Amino acid distributions in integral membrane protein structures. *Biochim Biophys Acta - Biomembr* **1512**, 1–14.
(30) Vascon, F. Protein Electrostatics: From Computational and Structural Analysis to Discovery of Functional Fingerprints and Biotechnological Design. *Comput Struct Biotechnol J* **2020**, *18*, 1774–1789.
(31) Uemura, E. Large-Scale Aggregation Analysis of Eukaryotic Proteins Reveals an Involvement of Intrinsically Disordered Regions in Protein Folding. *Sci Rep* **2018**, *8*, 678.
(32) Hebditch, M.; Warwicker, J. Web-Based Display of Protein Surface and pH-Dependent Properties for Assessing the Developability of Biotherapeutics. *Sci Rep* **1969**, *9*.
(33) Kim, S. Effect of Protein Surface Charge Distribution on Protein–Polyelectrolyte Complexation. *Biomacromolecules* **2020**, *21*, 3026–3037.
(34) McConkey, E. H. Molecular Evolution, Intracellular Organization, and the Quinary Structure of Proteins. *Proc. Natl. Acad. Sci.* **1982**, *79* (10), 3236–3240. https://doi.org/10.1073/pnas.79.10.3236.
(35) Rivas, G.; Minton, A. P. Macromolecular Crowding In Vitro , In Vivo , and In Between. *Trends Biochem. Sci.* **2016**, *41* (11), 970–981. https://doi.org/10.1016/j.tibs.2016.08.013.
(36) Guin, D.; Gruebele, M. Weak Chemical Interactions That Drive Protein Evolution: Crowding, Sticking, and Quinary Structure in Folding and Function. *Chem. Rev.* **2019**, *119* (18), 10691–10717. https://doi.org/10.1021/acs.chemrev.8b00753.
(37) Gimeno, A.; Valverde, P.; Ardá, A.; Jiménez-Barbero, J. Glycan Structures and Their Interactions with Proteins. A NMR View. *Curr. Opin. Struct. Biol.* **2020**, *62*, 22–30. https://doi.org/10.1016/j.sbi.2019.11.004.
(38) Savir, Y.; Tlusty, T. Conformational Proofreading: The Impact of Conformational Changes on the Specificity of Molecular Recognition. *PLoS One* **2007**, *2*, e468.
(39) Savir, Y.; Tlusty, T. The Ribosome as an Optimal Decoder: A Lesson in Molecular Recognition. *Cell* **2013**, *153* (2), 471–479. https://doi.org/10.1016/j.cell.2013.03.032.
(40) Savir, Y.; Tlusty, T. RecA-Mediated Homology Search as a Nearly Optimal Signal Detection System. *Mol Cell* **2010**, *40* (3), 388–396. https://doi.org/10.1016/j.molcel.2010.10.020.
(41) Fadeel, B. Hide and Seek: Nanomaterial Interactions With the Immune System. *Front. Immunol.* **2019**, *10*, 133. https://doi.org/10.3389/fimmu.2019.00133.

(42) Eisenberg, D.; Weiss, R. M.; Terwilliger, T. C. The Hydrophobic Moment Detects Periodicity in Protein Hydrophobicity. *Proc Natl Acad Sci* **1984**, *81*, 140–144.
(43) Barlow, D. J.; Thornton, J. M. The Distribution of Charged Groups in Proteins. *Biopolymers* **1986**, *25*, 1717–1733.
(44) Takashima, S.; Asami, K. Calculation and Measurement of the Dipole Moment of Small Proteins: Use of Protein Data Base. *Biopolymers* **1993**, *33*, 59–68.
(45) Antosiewicz, J. Computation of the Dipole Moments of Proteins. *Biophys J* **1995**, *69*, 1344–1354.
(46) Eisenhaber, F. Hydrophobic Regions on Protein Surfaces. Derivation of the Solvation Energy from Their Area Distribution in Crystallographic Protein Structures. *Protein Sci* **1996**, *5*, 1676–1686.
(47) Lijnzaad, P.; Berendsen, H. J. C.; Argos, P. Hydrophobic Patches on the Surfaces of Protein Structures. *Proteins Struct Funct Bioinforma* **1996**, *25*, 389–397.
(48) Chirgadze, Y. N.; Larionova, E. A. Spatial Sign-Alternating Charge Clusters in Globular Proteins. *Protein Eng Sel* **1999**, *12*, 101–105.
(49) Nicolau, D. V.; Paszek, E.; Fulga, F.; Nicolau, D. V. Protein Molecular Surface Mapped at Different Geometrical Resolutions. *PLoS One* **2013**, *8*.
(50) Nicolau, D. V.; Paszek, E.; Fulga, F.; Nicolau, D. V. Mapping Hydrophobicity on the Protein Molecular Surface at Atom-Level Resolution. *PLoS One* **2014**, *9*, 1–28.
(51) Wang, W.; Nema, S.; Teagarden, D. Protein Aggregation—Pathways and Influencing Factors. *Int J Pharm* **2010**, *390*, 89–99.
(52) Pillai, P. P.; Kowalczyk, B.; Pudlo, W. J.; Grzybowski, B. A. Electrostatic Titrations Reveal Surface Compositions of Mixed, On-Nanoparticle Monolayers Comprising Positively and Negatively Charged Ligands. *J Phys Chem C* **2016**, *120*, 4139–4144.
(53) Pillai, P. P.; Huda, S.; Kowalczyk, B.; Grzybowski, B. A. Controlled pH Stability and Adjustable Cellular Uptake of Mixed- Charge Nanoparticles. *J Am Chem Soc* **2013**, *2*, 1–4.
(54) Siek, M.; Kandere-Grzybowska, K.; Grzybowski, B. A. M.-C. pH-Responsive Nanoparticles for Selective Interactions with Cells, Organelles, and Bacteria. *Acc. Mater Res* **2020**, *1*, 188–200.
(55) Pillai, P. P.; Kowalczyk, B.; Grzybowski, B. A. Self-Assembly of like-Charged Nanoparticles into Microscopic Crystals. *Nanoscale* **2016**, *8*, 157–161.
(56) Pillai, P. P.; Kowalczyk, B.; Kandere-Grzybowska, K.; Borkowska, M.; Grzybowski, B. A. Engineering Gram Selectivity of Mixed-Charge Gold Nanoparticles by Tuning the Balance of Surface Charges. *Angew Chem. Int Ed* **2016**, *55*, 8610–8614.
(57) Borkowska, M. Targeted Crystallization of Mixed-Charge Nanoparticles in Lysosomes Induces Selective Death of Cancer Cells. *Nat Nanotechnol* **2020**, *15*, 331–341.
(58) Word, J. M.; Lovell, S. C.; Richardson, J. S.; Richardson, D. C. Asparagine and Glutamine: Using Hydrogen Atom Contacts in the Choice of Side-Chain Amide orientation11Edited by J. *Thornton J Mol Biol* **1999**, *285*, 1735–1747.
(59) Dolinsky, T. J. PDB2PQR: Expanding and Upgrading Automated Preparation of Biomolecular Structures for Molecular Simulations. *Nucleic Acids Res* **2007**, *35*, 522–525.
(60) Sanner, M. F.; Olson, A. J.; Spehner, J.-C. Reduced Surface: An Efficient Way to Compute Molecular Surfaces. *Biopolymers* **1996**, *38*, 305–320.
(61) Zhou, Q. PyMesh: Geometry Processing Library for Python, 2019. https//github.com/PyMesh/PyMesh.
(62) Jurrus, E. Improvements to the APBS Biomolecular Solvation Software Suite. *Protein Sci* **2018**, *27*, 112–128.
(63) Kortemme, T.; Morozov, A. V.; Baker, D. An Orientation-Dependent Hydrogen Bonding Potential Improves Prediction of Specificity and Structure for Proteins and Protein–Protein Complexes. *J Mol Biol* **2003**, *326*, 1239–1259.
(64) Kyte, J.; Doolittle, R. F. A Simple Method for Displaying the Hydropathic Character of a Protein. *J Mol Biol* **1982**, *157*, 105–132.
(65) Rego, N. B.; Xi, E.; Patel, A. J. Identifying Hydrophobic Protein Patches to Inform Protein Interaction Interfaces. *Proc Natl Acad Sci* **2021**, *118*.

(66) Qiao, B.; Jiménez-Ángeles, F.; Nguyen, T. D.; Cruz, M. Water Follows Polar and Nonpolar Protein Surface Domains. *Proc Natl Acad Sci* **2019**, *116*, 19274–19281.

(67) McBride, J. M.; Tlusty, T. Slowest-First Protein Translation Scheme: Structural Asymmetry and Co-Translational Folding. *Biophys. J.* **2021**, *120* (24), 5466–5477. https://doi.org/10.1016/j.bpj.2021.11.024.

(68) UniProt: The Universal Protein Knowledgebase in 2021. *Nucleic Acids Res* **2021**, *49*, 480–489.

(69) Klopfenstein, D. V.; Zhang, L.; Pedersen, B. S.; Ramírez, F.; Warwick Vesztrocy, A.; Naldi, A.; Mungall, C. J.; Yunes, J. M.; Botvinnik, O.; Weigel, M.; Dampier, W.; Dessimoz, C.; Flick, P.; Tang, H. GOATOOLS: A Python Library for Gene Ontology Analyses. *Sci. Rep.* **2018**, *8* (1), 10872. https://doi.org/10.1038/s41598-018-28948-z.

(70) Huerta-Cepas, J.; Serra, F.; Bork, P. ETE 3: Reconstruction, Analysis, and Visualization of Phylogenomic Data. *Mol. Biol. Evol.* **2016**, *33* (6), 1635–1638. https://doi.org/10.1093/molbev/msw046.

(71) Dawson-Haggerty, M. Trimesh, 2019.

(72) Dijkstra, E. W. A Note on Two Problems in Connexion with Graphs. *Numer Math* **1959**, *1*, 269–271.

(73) Allen, M. P.; Tildesley, D. J. *Computer Simulation of Liquids*, Second edition.; Oxford University Press: Oxford, 2017.

(74) Abraham, M. J.; Murtola, T.; Schulz, R.; Páll, S.; Smith, J. C.; Hess, B.; Lindahl, E. GROMACS: High Performance Molecular Simulations through Multi-Level Parallelism from Laptops to Supercomputers. *SoftwareX* **2015**, *1–2*, 19–25. https://doi.org/10.1016/j.softx.2015.06.001.

(75) Varadi, M.; Anyango, S.; Deshpande, M.; Nair, S.; Natassia, C.; Yordanova, G.; Yuan, D.; Stroe, O.; Wood, G.; Laydon, A.; Žídek, A.; Green, T.; Tunyasuvunakool, K.; Petersen, S.; Jumper, J.; Clancy, E.; Green, R.; Vora, A.; Lutfi, M.; Figurnov, M.; Cowie, A.; Hobbs, N.; Kohli, P.; Kleywegt, G.; Birney, E.; Hassabis, D.; Velankar, S. AlphaFold Protein Structure Database: Massively Expanding the Structural Coverage of Protein-Sequence Space with High-Accuracy Models. *Nucleic Acids Res.* **2022**, *50* (D1), D439–D444. https://doi.org/10.1093/nar/gkab1061.

(76) Zhuang, Q.; Yang, Z.; Sobolev, Y. I.; Beker, W.; Kong, J.; Grzybowski, B. A. Control and Switching of Charge-Selective Catalysis on Nanoparticles by Counterions. *ACS Catal.* **2018**, *8* (8), 7469–7474. https://doi.org/10.1021/acscatal.8b01323.

(77) Meineke, M. A.; Vardeman, C. F.; Lin, T.; Fennell, C. J.; Gezelter, J. D. OOPSE: An Object-Oriented Parallel Simulation Engine for Molecular Dynamics. *J. Comput. Chem.* **2005**, *26* (3), 252–271. https://doi.org/10.1002/jcc.20161.

(78) Cornell, W. D.; Cieplak, P.; Bayly, C. I.; Gould, I. R.; Merz, K. M.; Ferguson, D. M.; Spellmeyer, D. C.; Fox, T.; Caldwell, J. W.; Kollman, P. A. A Second Generation Force Field for the Simulation of Proteins, Nucleic Acids, and Organic Molecules. *J. Am. Chem. Soc.* **1995**, *117* (19), 5179–5197. https://doi.org/10.1021/ja00124a002.

(79) Heinz, H.; Vaia, R. A.; Farmer, B. L.; Naik, R. R. Accurate Simulation of Surfaces and Interfaces of Face-Centered Cubic Metals Using 12−6 and 9−6 Lennard-Jones Potentials. *J. Phys. Chem. C* **2008**, *112* (44), 17281–17290. https://doi.org/10.1021/jp801931d.

(80) Case, D. A.; Cheatham, T. E.; Darden, T.; Gohlke, H.; Luo, R.; Merz, K. M.; Onufriev, A.; Simmerling, C.; Wang, B.; Woods, R. J. The Amber Biomolecular Simulation Programs. *J. Comput. Chem.* **2005**, *26* (16), 1668–1688. https://doi.org/10.1002/jcc.20290.

(81) Moelbert, S.; Emberly, E.; Tang, C. Correlation between Sequence Hydrophobicity and Surface-Exposure Pattern of Database Proteins. *Protein Sci* **2004**, *13*, 752–762.

(82) Miller, S.; Janin, J.; Lesk, A. M.; Chothia, C. Interior and surface of monomeric proteins. *J Mol Biol* **1987**, *196*, 641–656.

(83) Treger, M.; Westhof, E. Statistical analysis of atomic contacts at RNA-protein interfaces. *J Mol Recognit* **2001**, *14*, 199–214.

(84) Bak, J. H. *Shaping the Information Channel:Molecules, Cells, and Experiments*; Princeton University, Princeton, 2016.

(85) Wade, R. C.; Gabdoulline, R. R.; Lüdemann, S. K.; Lounnas, V. Electrostatic Steering and Ionic Tethering in Enzyme–Ligand Binding: Insights from Simulations. *Proc Natl Acad Sci* **1998**, *95*, 5942–5949.

(86) Han, X.; Sit, A.; Christoffer, C.; Chen, S.; Kihara, D. A Global Map of the Protein Shape Universe. *PLOS Comput Biol* **2019**, *15*, 1–23.

(87) Taylor, R.; W., M. T.; J.; Turnell, G.; W. An Ellipsoidal Approximation of Protein Shape. *J Mol Graph* **1983**, *1*, 30–38.

(88) Rawat, N.; Biswas, P. Size, Shape, and Flexibility of Proteins and DNA. *J Chem Phys* **2009**, *131*, 165104.

(89) Hass, J.; Koehl, P. How round is a protein? Exploring protein structures for globularity using conformal mapping. *Front Mol Biosci* **2014**, *1*, 26.

(90) Dima, R. I.; Thirumalai, D. Asymmetry in the Shapes of Folded and Denatured States of Proteins. *J Phys Chem B* **2004**, *108*, 6564–6570.

(91) Shannon, G.; Marples, C. R.; Toofanny, R. D.; Williams, P. M. Evolutionary Drivers of Protein Shape. *Sci Rep* **2019**, *9*, 11873.

(92) Wang, D.; Nap, R. J.; Lagzi, I.; Kowalczyk, B.; Han, S.; Grzybowski, B. A.; Szleifer, I. How and Why Nanoparticle's Curvature Regulates the Apparent p$K_a$ of the Coating Ligands. *J. Am. Chem. Soc.* **2011**, *133* (7), 2192–2197. https://doi.org/10.1021/ja108154a.

(93) Zhang, X.; Huang, R.; Gopalakrishnan, S.; Cao-Milán, R.; Rotello, V. M. Bioorthogonal Nanozymes: Progress towards Therapeutic Applications. *Trends Chem.* **2019**, *1* (1), 90–98. https://doi.org/10.1016/j.trechm.2019.02.006.

(94) Ouyang, Y.; Biniuri, Y.; Fadeev, M.; Zhang, P.; Carmieli, R.; Vázquez-González, M.; Willner, I. Aptamer-Modified $Cu^{2+}$-Functionalized C-Dots: Versatile Means to Improve Nanozyme Activities-"Aptananozymes." *J. Am. Chem. Soc.* **2021**, *143* (30), 11510–11519. https://doi.org/10.1021/jacs.1c03939.

(95) Zhang, J.; Huang, Z.; Xie, Y.; Jiang, X. Modulating the Catalytic Activity of Gold Nanoparticles Using Amine-Terminated Ligands. *Chem. Sci.* **2022**, *13* (4), 1080–1087. https://doi.org/10.1039/D1SC05933E.

(96) Ji, X.; Li, Q.; Song, H.; Fan, C. Protein-Mimicking Nanoparticles in Biosystems. *Adv. Mater.* **2022**, *34* (37), 2201562. https://doi.org/10.1002/adma.202201562.

(97) Zangiabadi, M.; Zhao, Y. Synergistic Hydrolysis of Cellulose by a Blend of Cellulase-Mimicking Polymeric Nanoparticle Catalysts. *J. Am. Chem. Soc.* **2022**, *144* (37), 17110–17119. https://doi.org/10.1021/jacs.2c06848.

(98) Kim, M. On-Nanoparticle Gating Units Render an Ordinary Catalyst Substrate- And Site-Selective. *J Am Chem Soc* **2021**, *143*, 1807–1815.

(99) Ahumada, J. C.; Ahumada, G.; Sobolev, Y.; Kim, M.; Grzybowski, B. A. On-Nanoparticle Monolayers as a Solute-Specific, Solvent-like Phase. *Nanoscale* **2023**, *15* (13), 6379–6386. https://doi.org/10.1039/D2NR06341G.

(100) Jackson, A. M.; Myerson, J. W.; Stellacci, F. Spontaneous Assembly of Subnanometre-Ordered Domains in the Ligand Shell of Monolayer-Protected Nanoparticles. *Nat. Mater.* **2004**, *3* (5), 330–336. https://doi.org/10.1038/nmat1116.

(101) Singh, C.; Ghorai, P. K.; Horsch, M. A.; Jackson, A. M.; Larson, R. G.; Stellacci, F.; Glotzer, S. C. Entropy-Mediated Patterning of Surfactant-Coated Nanoparticles and Surfaces. *Phys. Rev. Lett.* **2007**, *99* (22), 226106. https://doi.org/10.1103/PhysRevLett.99.226106.

(102) Luo, Z.; Marson, D.; Ong, Q. K.; Loiudice, A.; Kohlbrecher, J.; Radulescu, A.; Krause-Heuer, A.; Darwish, T.; Balog, S.; Buonsanti, R.; Svergun, D. I.; Posocco, P.; Stellacci, F. Quantitative 3D Determination of Self-Assembled Structures on Nanoparticles Using Small Angle Neutron Scattering. *Nat. Commun.* **2018**, *9* (1), 1343. https://doi.org/10.1038/s41467-018-03699-7.

(103) Labbadia, J.; Morimoto, R. I. The Biology of Proteostasis in Aging and Disease. *Annu Rev Biochem* **2015**, *84*, 435–464.

(104) Kim, Y. E.; Hipp, M. S.; Bracher, A.; Hayer-Hartl, M.; Ulrich Hartl, F. Molecular Chaperone Functions in Protein Folding and Proteostasis. *Annu. Rev. Biochem.* **2013**, *82* (1), 323–355. https://doi.org/10.1146/annurev-biochem-060208-092442.
(105) Bishop, K. J. M.; Wilmer, C. E.; Soh, S.; Grzybowski, B. A. Nanoscale Forces and Their Uses in Self-Assembly. *Small* **2009**, *5*, 1600–1630.
(106) Valiokas, R.; Östblom, M.; Svedhem, S.; Svensson, S. C. T.; Liedberg, B. Thermal Stability of Self-Assembled Monolayers: Influence of Lateral Hydrogen Bonding. *J Phys Chem B* **2002**, *106*, 10401–10409.
(107) Liao, Y. C.; Sun, H.; Weeks, B. L. Measuring the Activation Energy of Thiol Desorption Using Lateral Force Microscopy. *Scanning* **2012**, *34*, 200–205.
(108) Aiertza, M. K.; Odriozola, I.; Cabañero, G.; Grande, H.-J.; Loinaz, I. Single-Chain Polymer Nanoparticles. *Cell. Mol. Life Sci.* **2012**, *69* (3), 337–346. https://doi.org/10.1007/s00018-011-0852-x.
(109) Awino, J. K.; Zhao, Y. Protein-Mimetic, Molecularly Imprinted Nanoparticles for Selective Binding of Bile Salt Derivatives in Water. *J. Am. Chem. Soc.* **2013**, *135* (34), 12552–12555. https://doi.org/10.1021/ja406089c.
(110) Zhang, N.; Zhang, N.; Xu, Y.; Li, Z.; Yan, C.; Mei, K.; Ding, M.; Ding, S.; Guan, P.; Qian, L.; Du, C.; Hu, X. Molecularly Imprinted Materials for Selective Biological Recognition. *Macromol. Rapid Commun.* **2019**, *40* (17), 1900096. https://doi.org/10.1002/marc.201900096.
(111) Haupt, K.; Medina Rangel, P. X.; Bui, B. T. S. Molecularly Imprinted Polymers: Antibody Mimics for Bioimaging and Therapy. *Chem. Rev.* **2020**, *120* (17), 9554–9582. https://doi.org/10.1021/acs.chemrev.0c00428.
(112) Barbee, M. H.; Wright, Z. M.; Allen, B. P.; Taylor, H. F.; Patteson, E. F.; Knight, A. S. Protein-Mimetic Self-Assembly with Synthetic Macromolecules. *Macromolecules* **2021**, *54* (8), 3585–3612. https://doi.org/10.1021/acs.macromol.0c02826.
(113) Combes, A.; Tang, K.-N.; Klymchenko, A. S.; Reisch, A. Protein-like Particles through Nanoprecipitation of Mixtures of Polymers of Opposite Charge. *J. Colloid Interface Sci.* **2022**, *607*, 1786–1795. https://doi.org/10.1016/j.jcis.2021.09.080.
(114) Eckmann, J.-P.; Rougemont, J.; Tlusty, T. Colloquium: Proteins: The Physics of Amorphous Evolving Matter. *Rev Mod Phys* **2019**, *91*, 031001.
(115) Dutta, S.; Eckmann, J.-P.; Libchaber, A.; Tlusty, T. Green Function of Correlated Genes in a Minimal Mechanical Model of Protein Evolution. *Proc Natl Acad Sci USA* **2018**, *115* (20), E4559–E4568.
(116) Mitchell, M. R.; Tlusty, T.; Leibler, S. Strain Analysis of Protein Structures and Low Dimensionality of Mechanical Allosteric Couplings. *Proc Natl Acad Sci USA* **2016**, *113* (40), E5847–E5855.
(117) Tlusty, T.; Libchaber, A.; Eckmann, J.-P. Physical Model of the Genotype-to-Phenotype Map of Proteins. *Phys Rev X* **2017**, *7* (2), 021037.
(118) Tlusty, T. Self-Referring DNA and Protein: A Remark on Physical and Geometrical Aspects. *Philos. Trans. R. Soc. Math. Phys. Eng. Sci.* **2016**, *374* (2063), 20150070. https://doi.org/10.1098/rsta.2015.0070.
(119) Eckmann, J.; Tlusty, T. Dimensional Reduction in Complex Living Systems: Where, Why, and How. *BioEssays* **2021**, *43* (9), 2100062. https://doi.org/10.1002/bies.202100062.
(120) Drummond, D. A.; Wilke, C. O. The Evolutionary Consequences of Erroneous Protein Synthesis. *Nat Rev Genet* **2009**, *10*, 715–724.
(121) Tlusty, T. A Colorful Origin for the Genetic Code: Information Theory, Statistical Mechanics and the Emergence of Molecular Codes. *Phys. Life Rev.* **2010**, *7* (3), 362–376. https://doi.org/10.1016/j.plrev.2010.06.002.
(122) Tlusty, T. A Model for the Emergence of the Genetic Code as a Transition in a Noisy Information Channel. *J. Theor. Biol.* **2007**, *249* (2), 331–342. https://doi.org/10.1016/j.jtbi.2007.07.029.

(123) Li, Y.; Lin, H.; Zhou, W.; Sun, L.; Samanta, D.; Mirkin, C. A. Corner-, Edge-, and Facet-Controlled Growth of Nanocrystals. *Sci. Adv.* **2021**, *7* (3), eabf1410. https://doi.org/10.1126/sciadv.abf1410.

(124) Kinnear, C.; Moore, T. L.; Rodriguez-Lorenzo, L.; Rothen-Rutishauser, B.; Petri-Fink, A. Form Follows Function: Nanoparticle Shape and Its Implications for Nanomedicine. *Chem. Rev.* **2017**, *117* (17), 11476–11521. https://doi.org/10.1021/acs.chemrev.7b00194.

(125) Ridelman, Y.; Singh, G.; Popovitz-Biro, R.; Wolf, S. G.; Das, S.; Klajn, R. Metallic Nanobowls by Galvanic Replacement Reaction on Heterodimeric Nanoparticles. *Small* **2012**, *8* (5), 654–660. https://doi.org/10.1002/smll.201101882.

(126) Li, Y.; Yin, D.; Lee, S. Y.; Lv, Y. Engineered Polymer Nanoparticles as Artificial Chaperones Facilitating the Selective Refolding of Denatured Enzymes. *Proc. Natl. Acad. Sci.* **2024**, *121* (19), e2403049121. https://doi.org/10.1073/pnas.2403049121.

(127) Mayer, M. P.; Rüdiger, S.; Bukau, B. Molecular Basis for Interactions of the DnaK Chaperone with Substrates. *Biol. Chem.* **2000**, *381* (9–10). https://doi.org/10.1515/BC.2000.109.

(128) Saio, T.; Guan, X.; Rossi, P.; Economou, A.; Kalodimos, C. G. Structural Basis for Protein Antiaggregation Activity of the Trigger Factor Chaperone. *Science* **2014**, *344* (6184), 1250494. https://doi.org/10.1126/science.1250494.

(129) Hayer-Hartl, M.; Bracher, A.; Hartl, F. U. The GroEL–GroES Chaperonin Machine: A Nano-Cage for Protein Folding. *Trends Biochem. Sci.* **2016**, *41* (1), 62–76. https://doi.org/10.1016/j.tibs.2015.07.009.

(130) Lundqvist, M.; Stigler, J.; Elia, G.; Lynch, I.; Cedervall, T.; Dawson, K. A. Nanoparticle Size and Surface Properties Determine the Protein Corona with Possible Implications for Biological Impacts. *Proc. Natl. Acad. Sci.* **2008**, *105* (38), 14265–14270. https://doi.org/10.1073/pnas.0805135105.

(131) Bilardo, R.; Traldi, F.; Vdovchenko, A.; Resmini, M. Influence of Surface Chemistry and Morphology of Nanoparticles on Protein Corona Formation. *WIREs Nanomedicine Nanobiotechnology* **2022**, *14* (4), e1788. https://doi.org/10.1002/wnan.1788.

(132) Fröhlich, E. The Role of Surface Charge in Cellular Uptake and Cytotoxicity of Medical Nanoparticles. *Int. J. Nanomedicine* **2012**, 5577. https://doi.org/10.2147/IJN.S36111.

(133) Moyano, D. F. Fabrication of Corona-Free Nanoparticles with Tunable Hydrophobicity. *ACS Nano* **2014**, *8*, 6748–6755.

(134) Velachi, V.; Bhandary, D.; Singh, J. K.; Cordeiro, M. N. D. S. Striped Gold Nanoparticles: New Insights from Molecular Dynamics Simulations. *J Chem Phys* **2016**, *144*, 244710.

(135) Harkness, K. M.; Balinski, A.; McLean, J. A.; Cliffel, D. E. Nanoscale Phase Segregation of Mixed Thiolates on Gold Nanoparticles. *Angew Chem. - Int Ed* **2011**, *50*, 10554–10559.

# Supporting Information for,

# "Statistical Survey of Chemical and Geometric Patterns on Protein Surfaces as a Blueprint for Protein-mimicking Nanoparticles"

**Clustering protein surface properties**

In order to cluster regions of a surface by a surface property, we need to first establish boundaries to convert the continuous surface properties into discrete categories. To keep things simple, we only use three categories for each property corresponding to low, medium, and high; *i.e.,* for surface potential these categories can be thought of respectively as negatively charged, neutral, and positively charged categories. The choice of boundaries is constrained by the need to maintain symmetry about zero, so that positive and negative categories are equivalent and opposite. We chose boundaries for each property that produced the most uniform distributions of low, medium and high categories: Gaussian curvature, (-0.1, 0.1); mean curvature, (-0.5, 0.5); hydrophobicity, (-2, 2); surface potential, (-5, 5).

**Dynamics Analysis**

We measure root-mean-square-fluctuations (RMSF) using simulated trajectories of proteins and NPs. We unwrap the atomic positions so that they are not divided by the periodic boundaries.[98] We then align conformations to a reference conformation using the Kabsch algorithm.[99] For proteins we align the entire protein to the first frame of the simulation trajectory. For NPs we align conformations just using the gold atoms, since these do not move. For each atom we measure the RMSF as

$$\mathrm{RMSF} = \frac{1}{N+1}\sqrt{\sum_i \left(r_i - \langle r_i \rangle\right)^2}$$

where *r* is the position of an atom in frame *i*, and *N* is the number of frames. To examine large rearrangements of ligands we measure the movement of ligands on NP surfaces by tracking the vector, $v$, between an atom at the tip of the ligand (MUA, C11; TMA, N) and the NP centroid. We measure the angle made between this vector in frame *i* and the first frame, $\theta_i = \cos^{-1}(v_i \cdot v_0)$. We then measure the root-mean-square-angular-fluctuations (RMSAF) as

$$\mathrm{RMSAF} = \frac{1}{N+1}\sqrt{\sum_i \theta_i^2}$$

**Radial distribution function**

We measure the ligand-ligand radial distribution function[58] by measuring the average distance between all atoms in ligand *i* and all atoms in ligand *j*, and counting ligand-ligand distances in a histogram of bin width 0.5 Å. Since the NP is finite, the histogram is only normalized by radial volume.

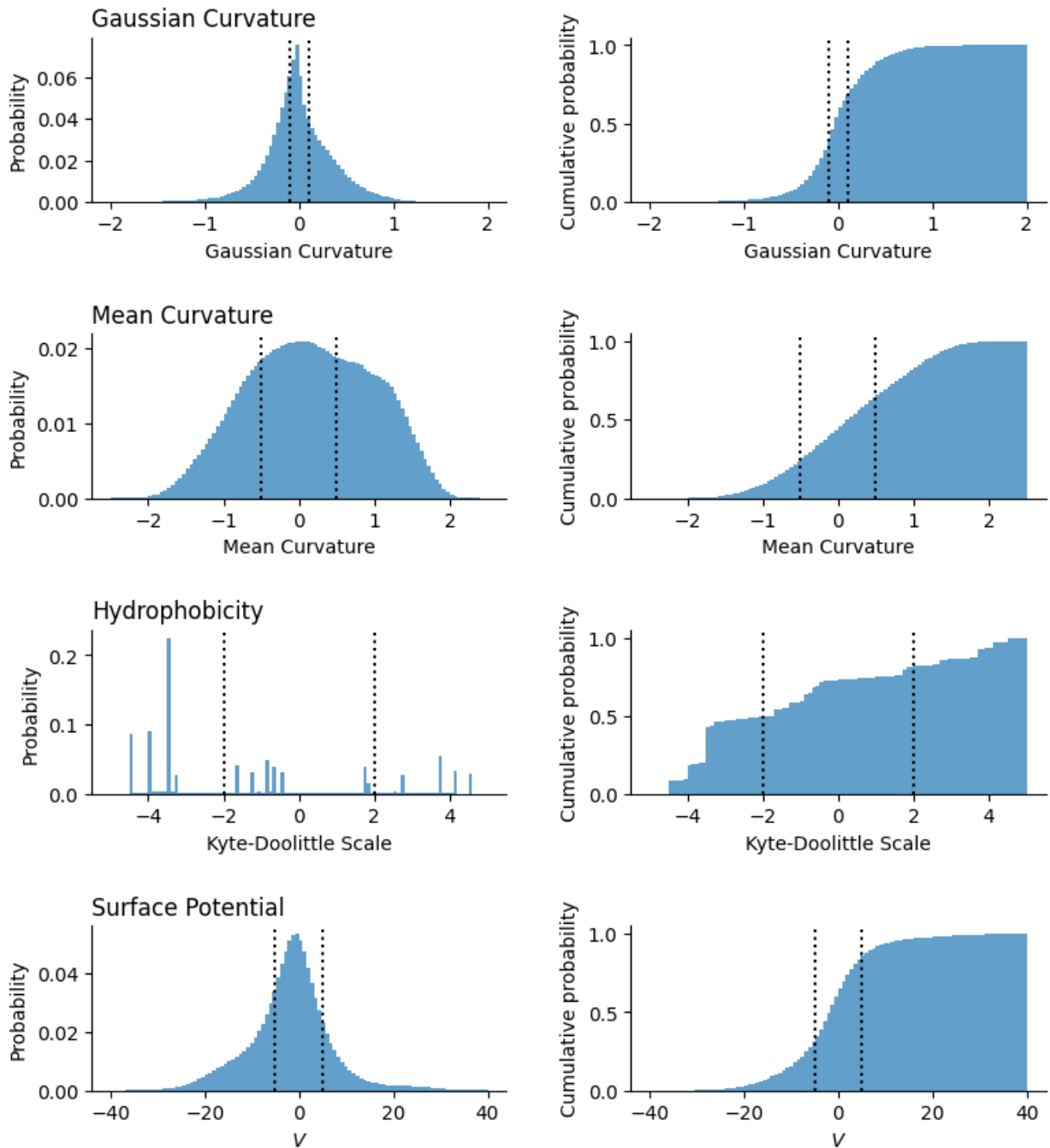

**SI Figure 1**

Distributions of surface properties across 14,963 proteins. Dotted lines indicate the boundaries used to separate surface properties into three types of clusters: low, medium, and high.

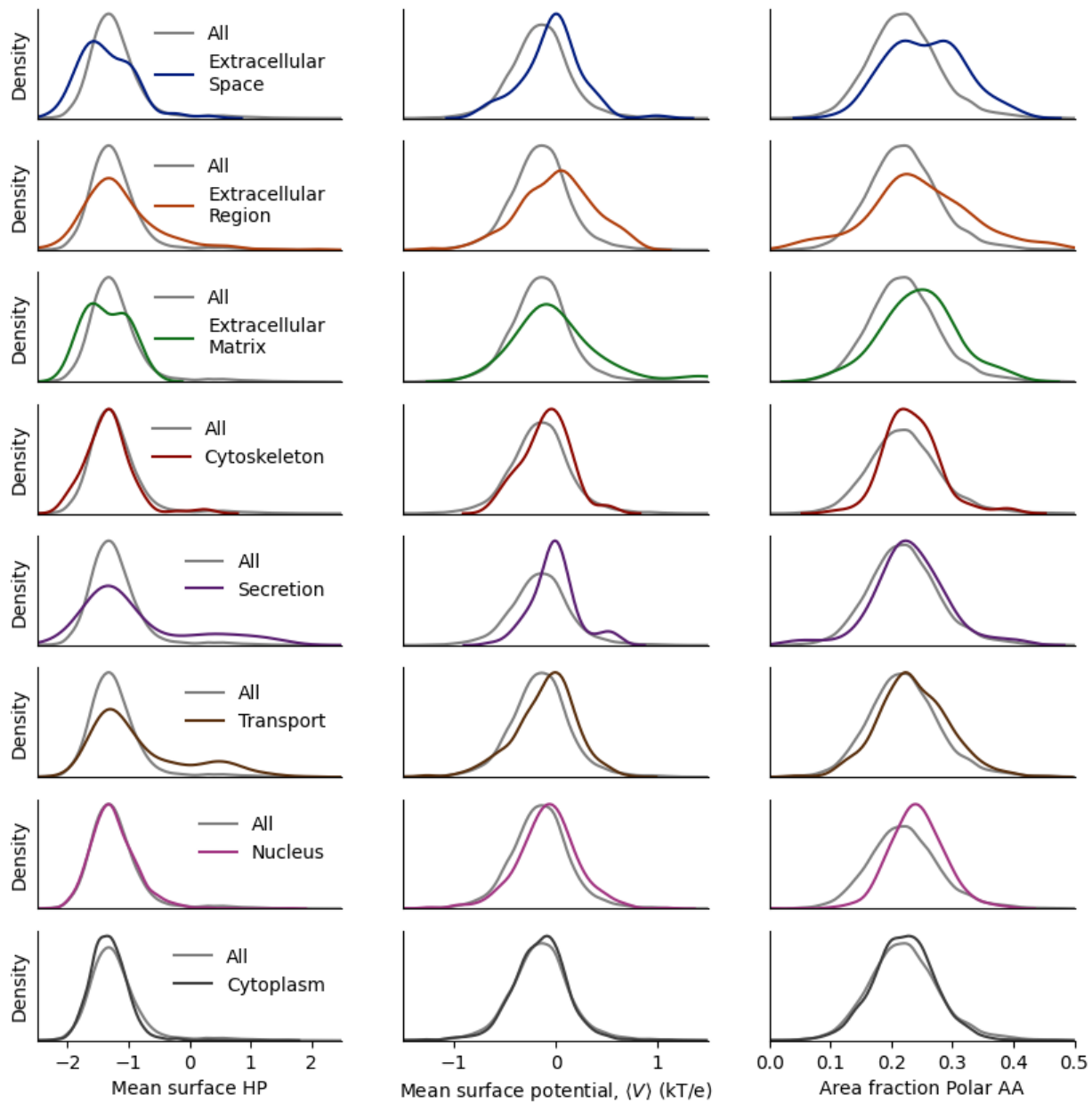

**SI Figure 2**

Distributions (kernel density estimates) of mean surface properties – hydrophobicity (HP), electric potential, and area fraction of polar residues – for all 14,963 proteins, and for groups of proteins separated by Gene Ontology terms: from top to bottom, GO:0005615, GO:0005576, GO:0031012, GO:0005856, GO:0046903, GO:0006810, GO:0005634, GO:0005737.

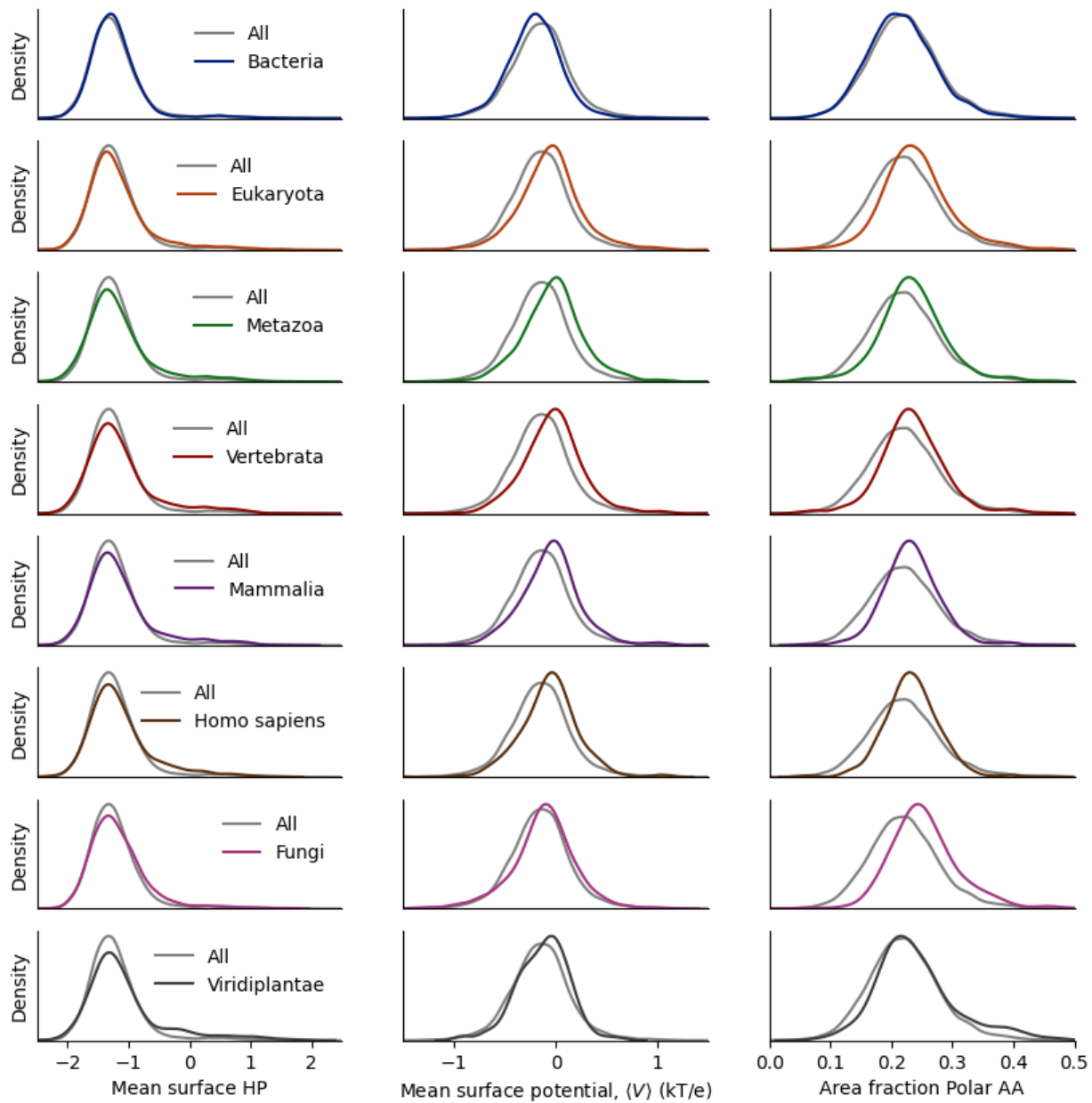

**SI Figure 3**

Distributions (kernel density estimates) of mean surface properties – hydrophobicity (HP), electric potential, and area fraction of polar residues – for all 14,963 proteins, and for groups of proteins separated by taxonomy: the NCBI taxonomy IDs, from top to bottom, are {2, 2759, 33208, 7742, 40674, 9606, 4751, 33090}.

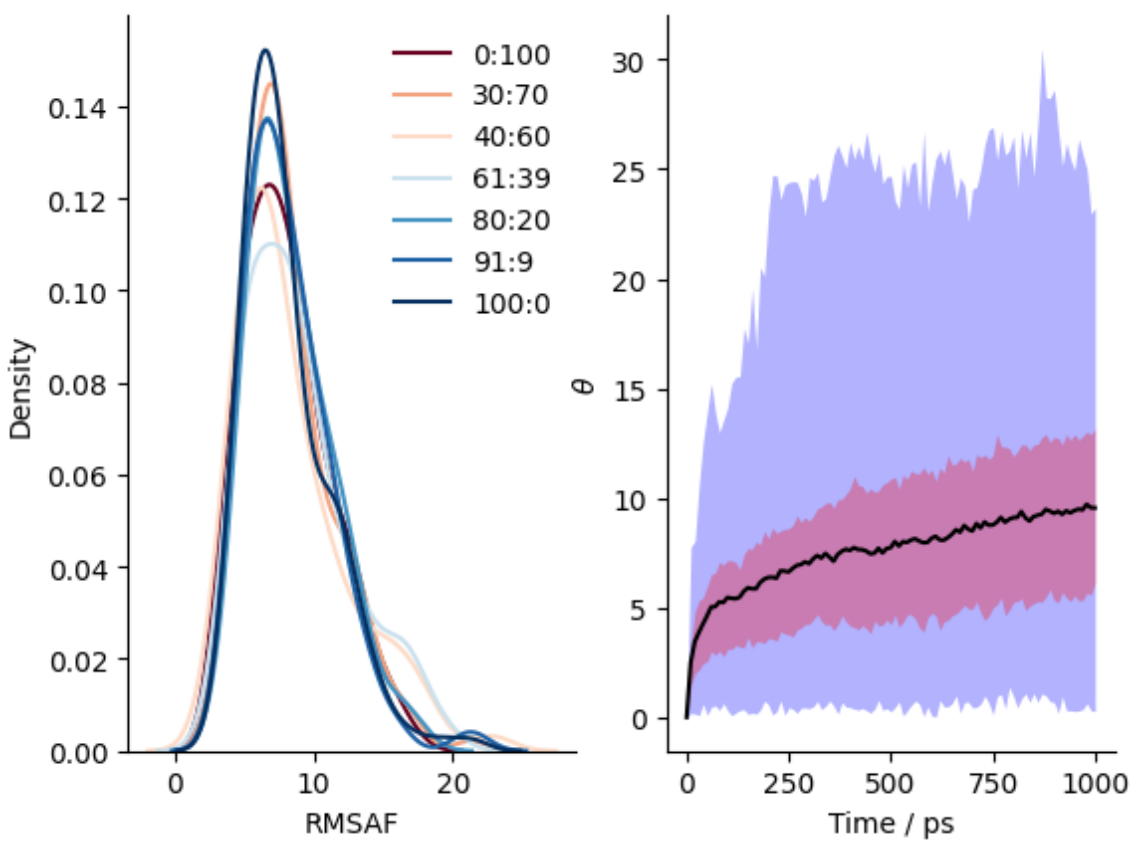

**SI Figure 4**

(Left) Distribution of RMSAF (root mean square angular fluctuation) per ligand for MCNPs. MCNPs of different ligand composition are colored according to TMA:MUA ligand ratio. (Right) Statistics of changes in the orientation of ligands w.r.t. the NP surface, $\theta$, with time: the average across all residues (black line); the interquartile range (pink area); the range (violet).

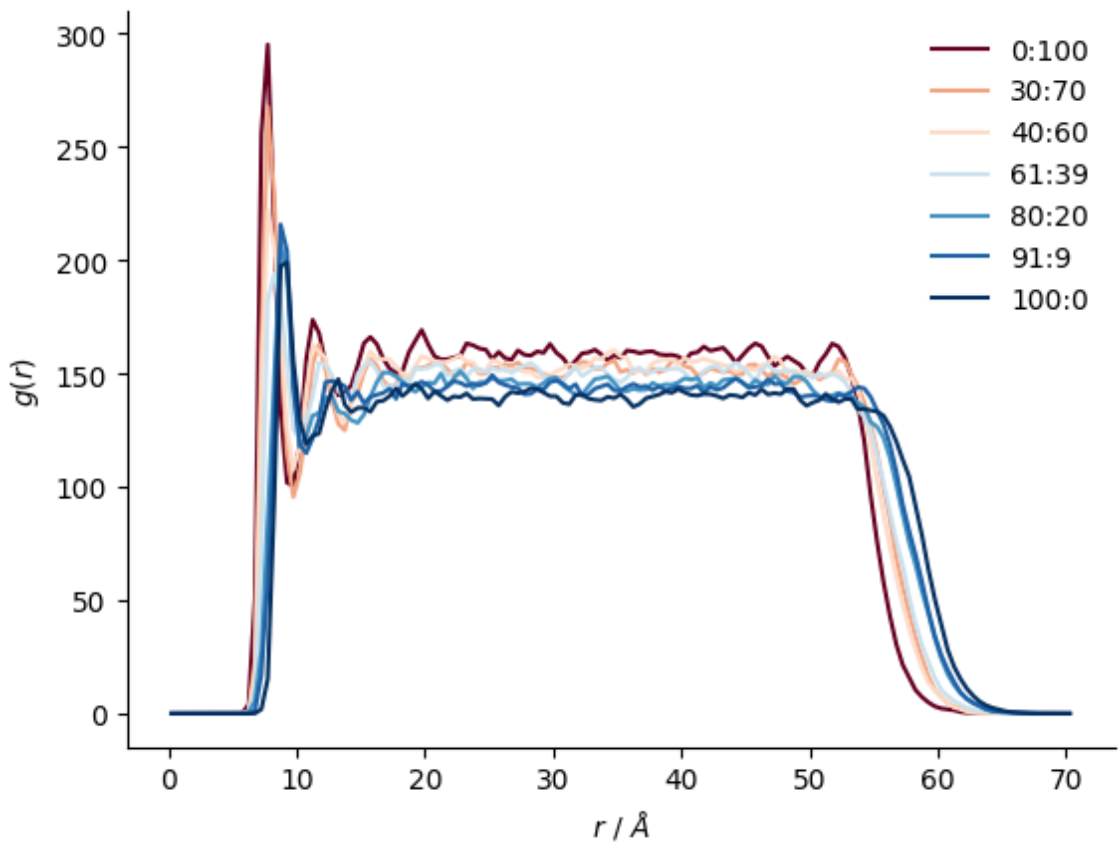

**SI Figure 5**

Ligand-ligand radial distribution function for MCNPs. MCNPs of different ligand composition are colored according to TMA:MUA ligand ratio.

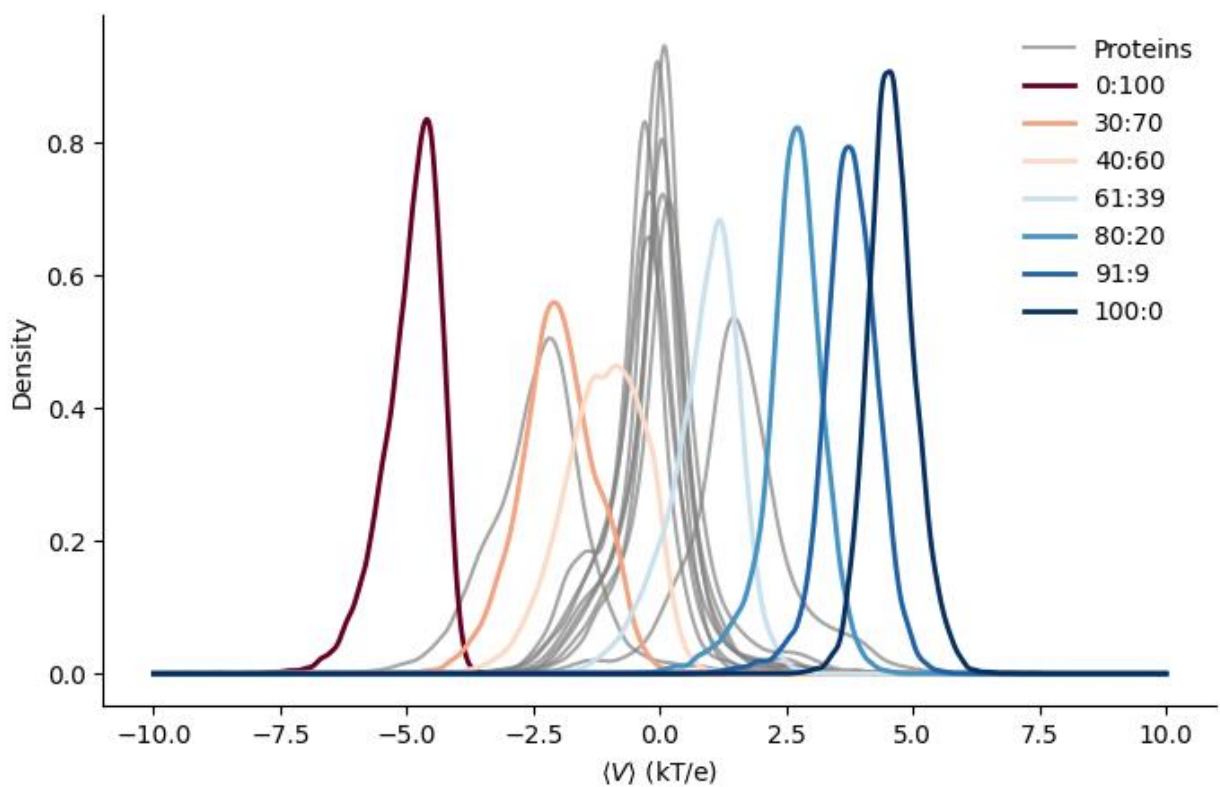

**SI Figure 6**

Distributions of electrostatic potential on the surface of NPs of different ligand composition (colored according to TMA:MUA ligand ratio) and proteins (grey). We selected 10 proteins to show differences in proteins with mean surface potential.

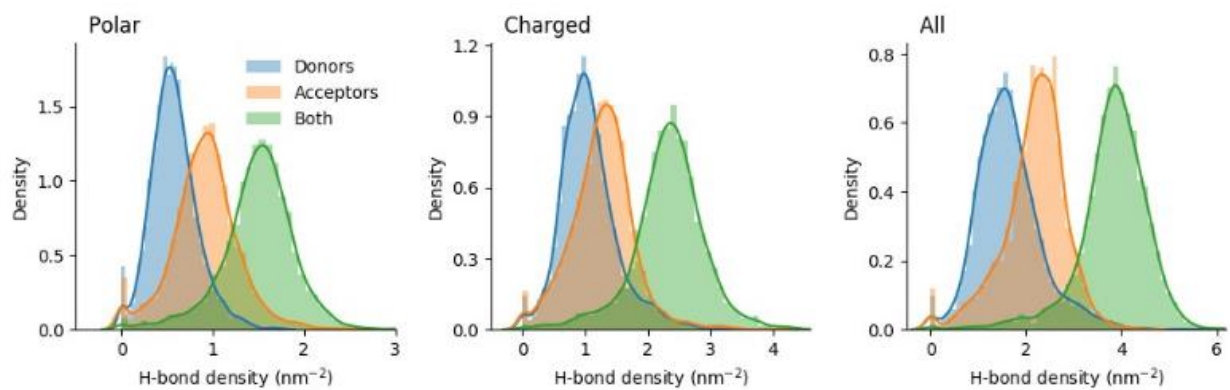

**SI Figure 7**

Distributions of hydrogen-bonding site density on protein surfaces. We show data respectively for: polar (neutral charge), charged (non-zero net charge) amino acids, and all amino acids; for donors, acceptors, and both types of sites.